\documentclass[showpacs,amsfonts,amssymb,amsmath]{iopart}
\usepackage{iopams}
\usepackage{cite}
\usepackage{graphicx,epsfig,color,bm,subfig,float}
\usepackage[applemac]{inputenc}

\newcommand{\eqs}[2]{equations~(\ref{#1}--\ref{#2})} 
\newcommand{\figs}[2]{figures~\ref{#1} and \ref{#2}} 

\newcommand{\bea}{\begin{eqnarray}}
\newcommand{\eea}{\end{eqnarray}}
\newcommand\be{\begin{equation}}
\newcommand\ee{\end{equation}}

\newcommand{\lang}{\left\langle}
\newcommand{\rang}{\right\rangle}
\renewcommand{\(}{\left(}
\renewcommand{\)}{\right)}
\renewcommand{\[}{\left[}
\renewcommand{\]}{\right]}
\newcommand{\lt}{\left}
\newcommand{\rt}{\right}

\newcommand{\p}{\partial}
\newcommand\tn{\tilde n}
\newcommand\tV{\tilde V}
\newcommand\tT{\tilde T}
\newcommand\vthi{v_{{\rm th}}}
\newcommand\kpar{k_\parallel}

\renewcommand{\etal}{\textit{et al.}}

\begin{document}


\title{Understanding the effect of sheared flow on microinstabilities}

\author{S L Newton$^1$, S C Cowley$^1$ and N F Loureiro$^2$} 

\address{$^1$ EURATOM/CCFE Fusion Association, Culham Science Centre, Abingdon, 
Oxon, OX14 3DB, UK}
\address{$^2$ Associa\c{c}\~ao EURATOM/IST, Instituto de Plasmas e Fus\~ao Nuclear -- Laborat\'orio Associado, Instituto Superior T\'ecnico, 1049-001 Lisboa, Portugal}

\ead{sarah.newton@ccfe.ac.uk}


\begin{abstract}

\noindent The competition between the drive and stabilization of plasma microinstabilities by sheared flow is 
investigated, focusing on the ion temperature gradient mode.
Using a twisting mode representation in sheared slab geometry, the characteristic equations have been 
formulated for a dissipative fluid model, developed rigorously from the gyrokinetic equation.
They clearly show that perpendicular flow shear convects perturbations along the field at a speed we denote by $Mc_s$ (where $c_s$ is the sound speed), 
whilst parallel flow shear enters as an instability driving term analogous to 
the usual temperature and density gradient effects.
For sufficiently strong perpendicular flow shear, $M >1$, the propagation of the system characteristics is 
unidirectional and no unstable eigenmodes may form. 
Perturbations are swept along the field, to be ultimately dissipated as they are sheared ever more strongly.
Numerical studies of the equations also reveal the existence of stable regions when $M < 1$, 
where the driving terms conflict.
However, in both cases transitory perturbations exist, which could attain substantial amplitudes before decaying. 
Indeed, for $M \gg 1$, they are shown to exponentiate $\sqrt{M}$ times. 
This may provide a subcritical route to turbulence in tokamaks.

\end{abstract}

\pacs{52.30.Gz, 52.35.Qz, 52.35.Ra}


\section{Introduction}
\label{secintro}

The confinement of heat and particles in tokamaks is determined by fine scale 
turbulence. 
This turbulence is driven by the equilibrium gradients in temperature, density 
and flow velocity~\cite{doyle}.
A central goal of fusion research is to understand and suppress this turbulence. 
There is increasing evidence that tokamak turbulence is strongly affected by 
sheared flows. 
For example, recent experiments on DIII-D show confinement increasing with 
toroidal flow shear~\cite{politzer}. 
Furthermore transport barriers, thin regions of the plasma with very large 
temperature gradients, have strongly sheared flow velocities~\cite{connor}.
It is presumed, but perhaps not proven, that turbulence and therefore heat 
diffusivity is suppressed by the sheared flows in the transport barriers. 
There is considerable literature on the stabilization of plasma instabilities 
by sheared flows (see for example~\cite{chen_90, hassam, waelbroeck1, artun_93, 
artun}). 
More generally it has been argued~\cite{biglari} that
flow shear suppresses turbulence by shearing apart the turbulent eddies.
There is also intense interest in the regulation of plasma turbulence by self-generated sheared flows --- see for 
example~\cite{wakatani, dimits, dorland, itosanddiamond} --- and the transport of plasma angular momentum, which is required in order to understand self-consistent flow profiles --- see for example~\cite{gurcan, peeters, smolyakov}.

Of course, not all shear flows are stabilizing. 
Indeed shear flows parallel to the magnetic field are destabilizing 
(this has been known since the 1960s, see\cite{dangelo, catto}). 
Also perpendicular shear flows can lower the instability threshold of 
gravitationally driven MHD instabilities --- see \cite{howes}. 
Typically strong flows in tokamaks are toroidal and are thus a mixture of 
parallel and perpendicular flow. What then is the optimum shear flow for 
confinement? Is increasing toroidal flow shear always desirable? 
In this paper we begin an examination of these questions by investigating 
the effect of flow shear on the ion temperature instability in a fluid 
slab model. 
Although more realistic descriptions have been investigated 
numerically~\cite{artun_93, artun, waltz, barnes, highcock}, this simple model 
allows a clear demonstration of some important features. 

The dynamics of the instability is investigated in coordinates that 
simultaneously shear with the flow in time and with the magnetic field in space. 
These transformations individually have a long history 
(see~\cite{kelvin, roberts}) and have been used in combination more 
recently~\cite{hameiri, waelchen, waltz2, howes}. 
Instabilities then take the form of twisting eddies that travel along the 
field at a speed $u_f$ (see figure~\ref{figure1} and equation~(\ref{u_f}) for a 
definition). 
We define a Mach number $M$ which is the ratio of $u_f$ to the sound speed.
Whilst others have used the more familiar Fourier 
representation~\cite{mattor, artun, waelbroeck1, rogister}, the ``twisting-shearing" representation used here clarifies aspects of the physics.
The fluid approximation analyzed here does not, of course, treat collisionless dissipation 
(Landau damping), nor does it accurately capture ion finite Larmor radius (FLR) 
effects and therefore cannot model accurately modern fusion devices. 
However it does provide a simple model to investigate the physics of sheared 
flow. 

We focus on the linear evolution of the ion temperature gradient mode (referred to as the ITG) 
and the mode driven by the parallel velocity gradient/shear (the PVG).
Note that, as has been discussed previously~\cite{mattor, waelbroeck1, rogister}, the PVG mode differs from the typical incompressible Kelvin Helmholtz instability, the free energy source being shear in the flow parallel to the magnetic field and compressibility effects being essential, allowing destabilization of sound waves.
We show that the twisting mode is localized where the perpendicular gradients 
are small so as to minimize the collisional dissipation. 
The exponential stability for moderate perpendicular flow shear ($M<1$) is 
investigated numerically.

At large sheared flow, $M>1$, exponential growth is stabilized as the 
perturbations are swept downstream. 
This result is proved in~\sref{secconvection} by considering the system 
characteristics. 
In an important paper, Waelbroeck \etal~\cite{waelbroeck1} showed 
(using the Fourier representation) that the parallel velocity gradient mode 
in a cold ion plasma is stabilized above a critical perpendicular flow shear. 
Their result can be translated into the criterion $M>1$ for their equations in 
the twisting shearing representation. 
A similar result was observed by Howes \etal~\cite{howes}, who showed that the 
MHD gravity-driven instability of a sheared slab is stabilized when $u_f$ 
exceeds the Alfv\'en speed.
 
It is well known that sustained turbulence is possible in systems without 
linear instability~\cite{orr, trefethen, grossmann}. 
Indeed it is a common cause of turbulence in fluid systems such as plane channel 
flow~\cite{orszag}. 
Waelbroeck \etal~\cite{waelbroeck2} suggested that transient perturbations driven 
by parallel flow shear could lead to turbulence above the critical perpendicular 
flow shear. 
We present in~\sref{sectransitory} an asymptotic calculation of transient 
growth in the large shear flow limit $M\gg 1$. 
This solution demonstrates that for $M\gg 1$ transient perturbations 
exponentiate many times ($\sqrt{M}$ times) before decaying. 
We have not considered the non-linear implications of these transient 
perturbations --- but one expects that they amplify enough to sustain turbulence. 
In fact recent toroidal gyro-kinetic simulations with sheared flow by 
Barnes \etal~\cite{barnes} and Highcock \etal~\cite{highcock} have identified regimes where exponential 
instabilities are stable but transient growth triggers turbulence. 

The analysis is presented as follows.
In~\sref{secsyseqns}, the collisional plasma response to an electrostatic 
perturbation in a magnetized slab is summarized, highlighting the distinct
 effects of parallel and perpendicular flow shear. 
The linear evolution of the system is considered for the remainder of the paper. 
The local dispersion relation (ignoring collisional dissipation) is discussed briefly in~\sref{secldr}, focussing on the ITG and PVG modes.
Then the effect of collisional dissipation and the localization by a sheared field without flow is considered in~\sref{secpardissip}. 
In~\sref{secconvection}, the evolution equations of the system are cast into characteristic form and the behaviour of transient perturbations analysed in the large shear flow limit, $M \gg 1$.
Results from a numerical study of the system are presented in~\sref{secnumerics}, supporting the analytical results developed and extending the investigation, particularly to moderate values of $M$.
Finally, a discussion of the implications of our results and a summary of the work are given in the conclusions,~\sref{conclusions}.


\section{System Equations}
\label{secsyseqns}

In this section we present the system of equations that we use to 
analyze the stability of ion temperature gradient and velocity gradient 
driven modes.


\subsection{Geometry}
\label{secgeometry}

\begin{figure}
\centering
\subfloat []{\includegraphics[width=0.47\textwidth]
{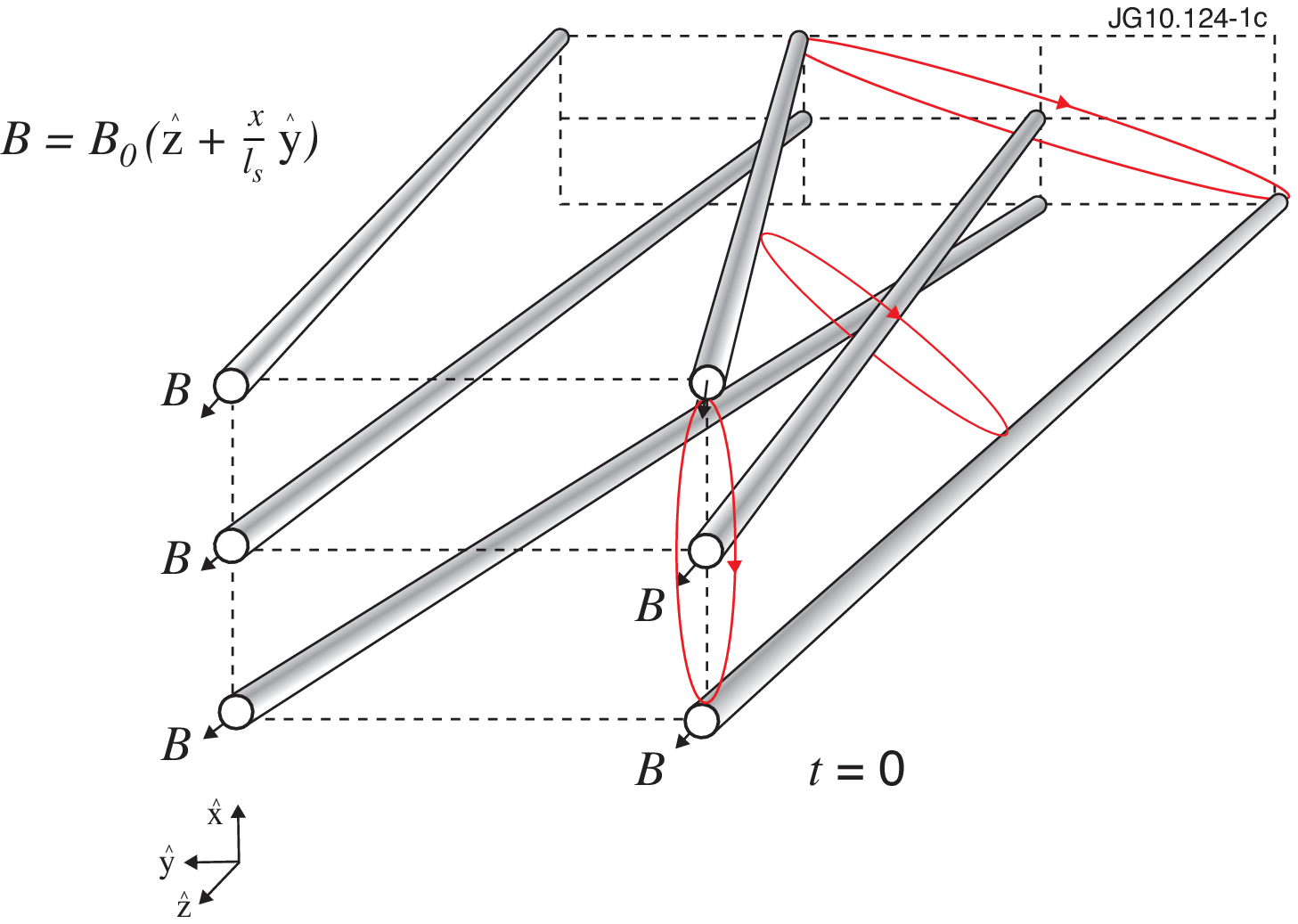}}
\subfloat []{\includegraphics[width=0.42\textwidth]{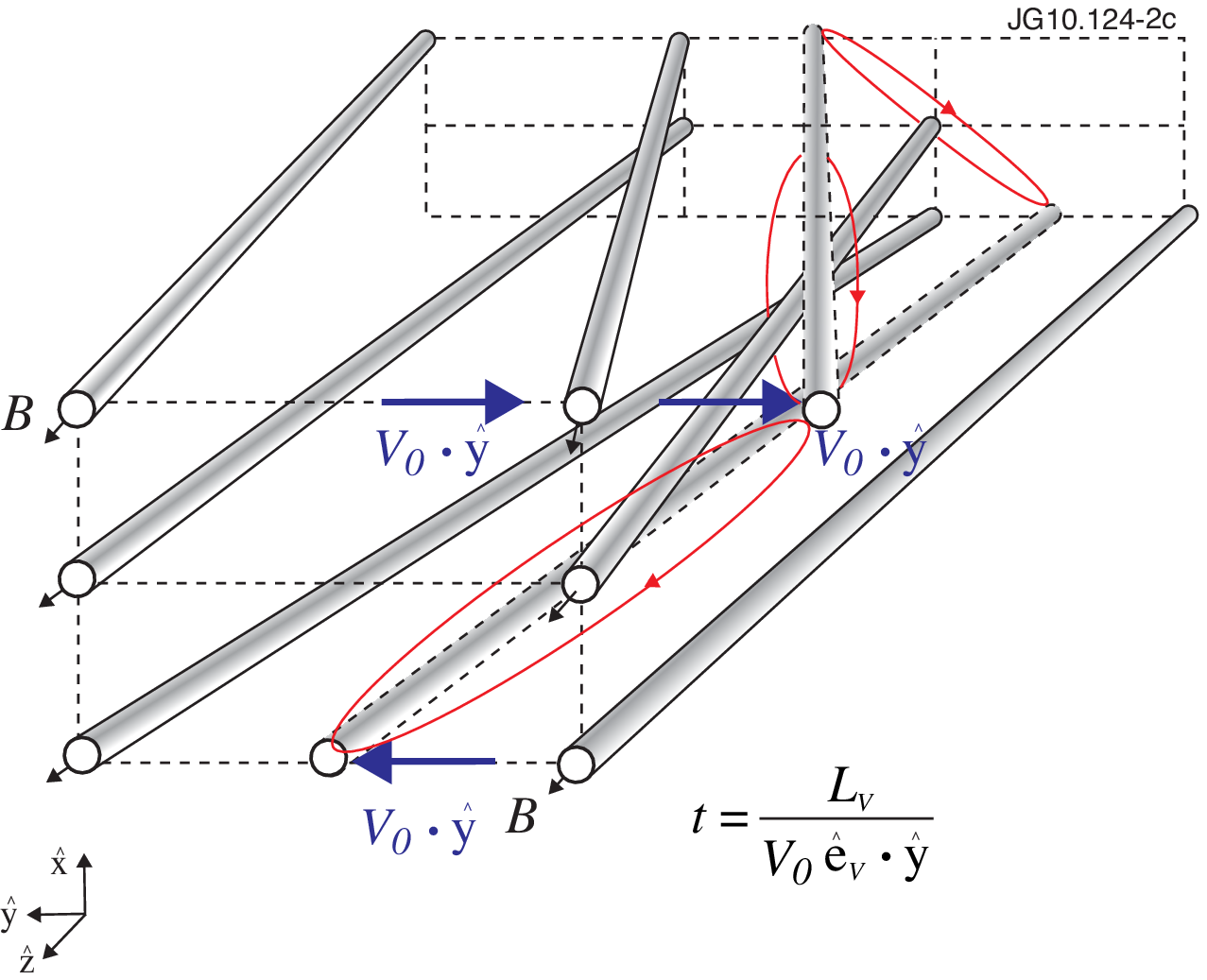}}
\caption{Geometry of the shearing magnetic field ${\bm B}$ and background flow ${\bm V}_0$. Looking in the $-{\bf \hat{z}}$ direction, the field is represented by flux tubes, twisting above and below the plane of $x=0$. Dashed grids are a guide, both $l_s$ and $L_v$ are taken to be negative. (a) Cross sections (red ovals) through a typical eddy at an initial time $t=0$. (b) Flux tubes are advected along $y$ by the perpendicular component of the sheared flow, see \eref{coords}. The eddy is thus twisted and its drive aligned position (eddy parallel to ${\bf \hat{x}}$) retreats along the field at speed $u_f = V_0 \(l_s/L_v\){\bf \hat e_v \cdot \hat y}$.}
\label{figure1}
\end{figure}

In order to highlight the main physical effects of sheared flow, we 
consider the stability of a simplified geometry --- the infinite plasma slab in the presence of sheared background magnetic and flow fields of the form:
\bea
\label{bshear}
\bm B &=& B_0\({\bf\hat z} + \frac{x}{l_s}{\bf\hat y}\),\\
\label{vshear}
\bm V_0 &=& V_0\frac{x}{L_v}{\bf \hat e_v},
\eea
where ${\bf \hat e_v}$ is a unit vector in the $(y,z)$-plane and $l_s,~L_v$ 
are, respectively, the characteristic scale length of the magnetic field and 
flow variation in the $x$-direction. This geometry is illustrated in \fref{figure1}, which is discussed further below. The variation of the background plasma density and temperature, which drives instability, is also taken to be in the $x$-direction (see \sref{secresponse}). This system has been analyzed many times, including studies of the 
stability of the ion temperature gradient driven mode with sheared background
flow~\cite{mattor,artun_93}. 
Note that we have chosen a frame in which there is no flow on the $x=0$ 
surface. The instabilities are assumed to be localized where 
$(x/L_v,~x/l_s) \ll 1$ so \eref{bshear} and \eref{vshear} can be thought of as
Taylor expansions about $x=0$ of more general field and flow profiles.
The sheared flow has components parallel and perpendicular to the magnetic field.
As we show, these two components produce distinctly different physical effects.

To analyze the system, it is most useful to implement a double-shearing 
coordinate transformation, which was outlined for a slab in~\cite{howes}. 
Such a transformation was introduced into the torus by Cooper~\cite{cooper} 
(see also~\cite{hameiri, waelchen, waltz}). 
This removes the complications of both the magnetic field shear and the time 
dependence associated with the equilibrium flow. 
It also aligns coordinate lines with the characteristics of the plasma 
response (the guiding centre motion and sound waves) and therefore simplifies both the analysis and the interpretation. In order to treat the effect of magnetic field shear, we first make 
the transformation~\cite{roberts} to the set of field line following coordinates:
\be
z^f = z, \qquad
y^f = y - \frac{x}{l_s} z, \qquad
x^f = x,
\label{fieldcoords}
\ee
with the property that:
\be
\bm B \cdot \nabla x^f = \bm B \cdot \nabla y^f = 0, \qquad
\bm B \cdot \nabla \equiv B_0 \frac{\p }{\p{z^f}}.
\ee

The component of the equilibrium flow perpendicular to the magnetic field 
convects (see \fref{figure1}(b)) the magnetic field structure, the guiding centres and the sound waves.
The time dependent transformation 
\be
y^* = y - V_0 \frac{x}{L_v} \({\bf \hat e_v \cdot \hat y}\) t
\ee
follows the shearing arising from the $y$ (perpendicular) flow component.
This type of transformation was first introduced by Kelvin~\cite{kelvin} to 
analyze instabilities due to sheared flow in fluids. 
Combining the two transformations yields the final coordinate 
transformation~\cite{howes}:
\be
t^{\prime} = t, \qquad
z^{\prime} = z + u_f t, \qquad
y^{\prime} = y - \frac{x}{l_s} z^{\prime}, \qquad
x^{\prime} = x,
\label{coords}
\ee
where we define the velocity:
\be
u_f = V_0 \frac{l_s}{L_v}{\bf \hat e_v \cdot \hat y}.
\label{u_f}
\ee
Thus we have transformed to a moving frame travelling along the $z$ direction 
(essentially the field lines) at velocity $-u_f$. 
In this frame the combined transformation is identical to \eref{fieldcoords} 
and therefore independent of time. 
The coordinates yield the simplifications:
\bea
\bm B \cdot \nabla x^{\prime} = \bm B \cdot \nabla y^{\prime} = 0, \qquad
\bm B \cdot \nabla \equiv B_0 \frac{\p }{\p{z^{\prime}}}, \\
\frac{\p }{\p t} + {\bm V_0\cdot{\bf \hat y}}\frac{\p }{\p y} = 
\frac{\p }{\p t^{\prime}} +  u_f\frac{\p }{\p z^{\prime}}. 
\eea
Note that in the moving frame the plasma flows along the field line at 
velocity $u_f$. 
\captionsetup[subfloat]{position=top}
\captionsetup[subfloat]{position=bottom}
Perturbations in this system take the form of eddies which are localized in $x$ and extended along the field line --- cross sections through a single eddy (red ovals) are shown at various $z$ positions in \fref{figure1}(a). The eddy tilts to remain on a surface of constant $y^\prime$, along which the plasma has its characteristic response. With time this surface, and hence the eddy, is twisted by the perpendicular component of the sheared flow, see \eref{coords}. This is shown in \fref{figure1}(b) and the drive aligned position of the eddy (eddy lies parallel to ${\bf \hat{x}}$) retreats along the field at the speed $u_f$. For sufficiently large $u_f$ the eddy dynamics is too slow to allow this response and the mode structure indicated becomes transitory (see \sref{secconvection}).


\subsection{Plasma Response}
\label{secresponse}

The equations used to describe the plasma response will now be presented. 
The plasma is taken to be quasineutral, consisting of electrons and a 
single (hydrogenic) ion species, of charge $e$ and mass $m$. 
The equilibrium density of both species is $n_0$. 
Only electrostatic perturbations are considered, with the perturbation frequency, 
$\omega$, much less than the ion cyclotron frequency, $\Omega = eB_0/m$. 
For simplicity, the equilibrium ion and electron temperatures are taken to be 
equal in the final form of the equations and denoted by $T$.  
We consider instabilities with phase velocities comparable to the ion 
thermal speed and therefore much slower than the electrons. 
The electrons thus have time to set up thermal equilibrium and so are taken 
to have an isothermal Boltzmann response to the perturbing potential $\phi$, 
giving the perturbed electron density
\be
\delta n_e = \frac{e\phi}{T}n_0.
\label{Boltzmann_response}
\ee

The ions are described by collisional fluid equations; the full details of their 
derivation are given in the appendix. 
The equations are derived from the gyro-kinetic equation~\cite{frieman,sugama} thus they embody the usual gyro-kinetic ordering: 
\be
\frac{\omega}{\Omega} \sim \mathcal O\(\frac{k_\parallel}{k_\perp}\) \sim 
\mathcal O\(\frac{\delta f}{F_0}\)\ll 1,
\label{gkorder}
\ee
as a primary expansion. 
Parallel (or perpendicular) is taken with respect to the equilibrium 
magnetic field, $k$ is the wavenumber of the perturbation and $\delta f$ is 
the perturbed distribution function, where $F_0$ represents the bulk distribution. We make a subsidiary collisional expansion, as shown in the appendix, with:
\be
\nu~\gg~\omega,~\omega^*,~\vthi k_\parallel,~u_f k_\parallel,~\nu k^2 \rho^2. 
\label{suborder}
\ee
Here $\nu$ is the ion-ion collision frequency, $\omega^*$ represents the drift 
frequencies associated with the background gradients (see \sref{seclinsys}), 
$\vthi=\(2T/m\)^{1/2}$ is the ion thermal velocity and $\rho = \vthi/\Omega$, 
is the ion gyroradius. 
All quantities on the right hand side of \eref{suborder} are treated as 
the same order in the derivation of the fluid equations.  
Note that $k\rho \sim \mathcal O \(\sqrt{\omega/\nu}\)\ll 1$.

Upon expanding the gyro-kinetic equation with respect to $\omega/\nu$, we 
obtain to lowest order a perturbed Maxwellian distribution (see appendix).  
The perturbed density, $\delta n$, parallel velocity, $\delta V_\parallel$, 
and temperature, $\delta T$, of this Maxwellian obey fluid equations --- 
the moment equations. 
The three moment equations for particle, parallel momentum and energy 
conservation may be combined with the requirement of quasineutrality on the 
perturbed electron and ion densities. 
This gives a closed set of equations for the evolution of the perturbed variables.
We employ the following normalizations, which are used in many gyro-kinetic 
codes (see for example~\cite{GS2}). They indicate the typical features associated 
with the unstable modes: the rapid perpendicular and long parallel spatial 
dependence, the characteristic acoustic timescale and the amplitude scaling.
\bea
x^{\prime}= \rho_s \tilde x, \qquad
& y^{\prime}=\rho_s \tilde y, \qquad
& z^{\prime}= l_s \tilde z, \qquad
t = \frac{l_s}{c_s} \tilde t, \\
\tV = \frac{\delta V_\parallel}{c_s}\frac{l_s}{\rho_s}, \qquad
& \tT = \frac{\delta T}{T}\frac{l_s}{\rho_s}, \qquad
& \tn = \frac{\delta n}{n_0}\frac{l_s}{\rho_s},
\eea
where $\rho_s = c_s/\Omega$ is the sound Larmor radius associated with the 
sound speed $c_s = \sqrt{\(\gamma_e + \gamma_i\)T/m}$. 
In this collisional model,  $\gamma_e = 1$ (isothermal) and $\gamma_ i = 5/3$ 
(adiabatic). 
The background gradients provide the instability drive. 
The scale-lengths for the equilibrium density, parallel velocity, temperature 
and the Mach number associated with the moving frame are:
\bea
\frac{1}{l_n} = \frac{d}{dx}\ln n_0, \qquad
& \frac{1}{l_T} = \frac{d}{dx}\ln T, \nonumber \\
\frac{1}{l_v} = \frac{1}{L_v}\frac{V_0}{c_s} {\bf \hat e_v} \cdot {\bf \hat{z}}, \qquad
& M = \frac{u_f}{c_s}.
\eea
The tildes denoting the final transformations will henceforth be dropped 
for convenience.

The set of three equations describing the evolution of a perturbation of the 
system in the doubly sheared system of coordinates thus takes the form:
\bea
\label{density} 
 \fl \(\frac{\p }{\p t} + M\frac{\p }{\p z}\)n
+\frac{\p V}{\p z} = \frac{3}{8}\frac{l_s}{l_n}\frac{\p n}{\p y}, \\
\label{paraflow} 
 \fl \(\frac{\p }{\p t} + M\frac{\p }{\p z}\)V
+\frac{3}{8}\frac{\p}{\p z}\(2 n+ T\) + \frac{3}{8}\left\{n , V \right\} =
\frac{3}{8}\frac{l_s}{l_v}\frac{\p n}{\p y}
+ \nu_\perp\nabla_\perp^2 V,\\
\label{temperature}
\fl \(\frac{\p }{\p t} + M\frac{\p }{\p z}\)T
+\frac{2}{3}\frac{\p V}{\p z} + \frac{3}{8}\left\{n,T\right\}
=\frac{3}{8}\frac{l_s}{l_T}\frac{\p n}{\p y}
+ \chi_\perp\nabla_\perp^2 T,
\eea
where
\be
\nabla_\perp^2 = \frac{\p^2}{\p y^2} + \(\frac{\p}{\p x} - z\frac{\p}{\p y}\)^2
\ee
and with the unit vector defined as ${\bf b} = {\bm B}/B$, for two arbitrary 
functions $p$ and $q$ the normalized Poisson bracket form is:
\bea
\left\{p, q \right\} = \( \nabla p \times \nabla q \)\cdot {\bf b} = \frac{\p p}{\p x}\frac{\p q}{\p y} - \frac{\p p}{\p y}\frac{\p q}{\p x}.
\eea
The time derivative appearing in these equations has a convective form: $\frac{\p}{\p{t}} + M \frac{\p}{\p{z}}$ --- in the moving frame the plasma is 
flowing along the field lines at a normalized velocity $M$. 
As discussed above, the potential $\phi$ is
proportional to $n$ (see \eref{Boltzmann_response}), so the perturbed $E \times B$ drift in our 
normalized time and spatial coordinates is given by $\nabla n \times{\bf b}$. 
Thus, nonlinear terms describe convection of the perturbations by the perturbed 
$E \times B$ drift. Since $\left\{n, n \right\}=0$, there is no nonlinearity in 
the density evolution equation. 
The instability drive comes from the first term on the right hand side of 
each of~\eqs{density}{temperature}. They are, respectively, the convection of 
the equilibrium density, parallel velocity and temperature along their gradients 
by the perturbed $E \times B$ drift. 
Dissipation results from the diffusive viscosity and conductivity terms. 
These are derived in the appendix and have the following normalized forms:
\be
\label{nu_chi_perp}
\left(\nu_\perp,\chi_\perp\right) = \left(\frac{9}{40},\frac{1}{4}\right)
\sqrt{\frac{2}{3}}\frac{l_s n_0 e^4 \ln \Lambda}{8 \pi^{3/2} \epsilon_0^2T^2 },
\ee
giving a Prandtl number of:
\be
P_r = \frac{\nu_\perp}{\chi_\perp} = \frac{9}{10}.
\ee
In the sheared coordinates the dissipative term becomes large at large $z$ --- 
this is just the mathematical manifestation of the shearing of the (field) 
coordinate lines. 
However, as we shall see, the perturbations follow the (field) coordinate 
lines and therefore the effective dissipation does indeed increase along field 
lines. 
As ion-ion collisions conserve ion momentum and therefore do not produce particle 
transport, there is no particle diffusion in the density equation. 
If we had included ion-electron collisions we would have obtained a particle diffusion of order: $D_\perp \sim \sqrt{m_e/m_i} ~ \chi_\perp$.
Note also the limitations pointed out in the introduction: a fluid model such as this cannot capture collisionless dissipation or FLR effects. However, its particular strength lies in the clear representation of the physical effects of sheared flow which it allows.

The background flow shear enters the system of~\eqs{density}{temperature} in two forms. 
First, as a parallel (along $z$) convective velocity, $M$, due to the 
perpendicular component of the flow shear, which acts on all the perturbed 
variables. 
Second as an instability drive, $\propto ~ l_s/l_v$, from the parallel component 
of the flow shear, which only acts on the parallel velocity perturbation. 
The magnitudes of these flow effects are linked, as both are proportional to 
the magnitude, $V_0$, of the imposed background flow, so:
\be
\frac{M}{\(l_s/l_v\)} = \frac{\bf \hat e_v \cdot \hat y}
{\bf \hat e_v \cdot \hat z} \equiv \tan{\theta_v}.
\label{vangle}
\ee
The aim of this paper is to understand the interaction of these two effects 
and whether sufficient flow shear can stabilize the system via strong convection, 
which will be seen to sweep the perturbations along the field lines into the 
dissipative regions at large $z$.


\subsection{Linearized System}
\label{seclinsys}

For the remainder of this paper we neglect the nonlinear terms in the equation 
set above and restrict ourselves to the investigation of the linear evolution 
of the system. 
All fields are taken to vary as $\exp(i k y)$ multiplied by a function of $z$. 
The perpendicular gradient operator then becomes:
 $\nabla_\perp^2 = - k^2 \( 1 + z^2\)$ and we define: $\nu_k=k^2 \nu_\perp$ and 
$\chi_k=k^2 \chi_\perp$. The linearized set of
equations then takes the form:
\bea
\(\frac{\p }{\p t} + M\frac{\p }{\p z}\)n
+\frac{\p V}{\p z}=i\omega_n^*n,
\label{neq} \\
\(\frac{\p }{\p t}+M\frac{\p }{\p z}\)V
+\frac{3}{8}\frac{\p}{\p z}\(2 n+ T\)=i\omega_v^*n
-\nu_k\(1+z^2\)V, \\
\(\frac{\p }{\p t}+M\frac{\p }{\p z}\)T
+\frac{2}{3}\frac{\p V}{\p z}=i\omega_T^*n
-\chi_k\(1+z^2\)T.
\label{teq}
\eea
The following effective drift frequencies have been introduced, 
characterizing the three background driving gradients:
\be
\omega_n^*=\frac{3k}{8}\frac{l_s}{l_n}, \qquad
\omega_v^*=\frac{3k}{8}\frac{l_s}{l_v}, \qquad
\omega_T^*=\frac{3k}{8}\frac{l_s}{l_T}.
\ee
For a perturbation of the form $\exp(i k y + ik_x x)$ the equations are identical if we make the transformation $z \rightarrow (z - k_x/k)$. As noted above the dissipative terms increase strongly with distance along the field line. 

In the following two sections analytical limits of this third order 
system are considered, which highlight the effects of parallel and perpendicular 
flow shear on the system.  We consider the local dispersion relation first.


\section{Local Dispersion Relation}
\label{secldr}

The dynamics of the basic modes present in this third-order system may be 
identified by considering the local, dissipationless form of the above equations. 
By neglecting the effects of dissipation, $\nu_k=0,~\chi_k=0$, the terms 
explicitly proportional to $z$ vanish, so we may look for solutions in the 
form $\exp(-i\omega t+i\kpar z)$. 
Noting the purely convective effect of the perpendicular flow shear, 
$M$, we define the frequency in the laboratory frame as:
\be
\omega^\prime = \omega - k_\parallel M.
\ee
The dispersion relation is then:
\be
\omega'^2 \(\omega'+\omega_n^*\) - \omega' \kpar \( \kpar - \omega_v^*\) + 
\frac{\kpar^2}{4} \( \frac{3}{2}\omega_T^* - \omega_n^*\) = 0.
\label{localdr}
\ee

When $\omega_n^*=\omega_v^*=0$ we recover the usual cubic dispersion relation for 
the collisional ion temperature gradient (ITG) instability~\cite{cowley}. 
The same limit for instability, $\omega_T^* > \(16/9\sqrt{3}\) k_\parallel$, 
applies here with finite $M$, but the corresponding oscillatory frequency of the 
mode has changed, by $-k_\parallel M$. 
For $\omega_T^* \gg k_\parallel$, \eref{localdr} reduces to the original ITG 
dispersion relation: $\omega'^3 = - 3\kpar^2\omega_T^*/8$, derived by 
Rudakov and Sagdeev \cite{rudakov-temp}.

In the limit $\omega_n^*=\omega_T^*=0$ the dispersion relation simplifies to a 
quadratic:
\be
\omega'^2 = \kpar \( \kpar - \omega_v^* \).
\label{wvmodedr}
\ee
This describes the parallel velocity gradient (PVG) instability introduced by 
previous authors~\cite{catto, waelbroeck1}, with instability if:
\be
\omega_v^* > k_\parallel.
\label{instcondition}
\ee
The most unstable mode has $k^{max}_\parallel = \omega_v^*/2$ and thus a growth 
rate $\gamma_{max} = \omega_v^*/2$.
The stability of the $\omega_v^*$-driven mode is seen from \eref{instcondition} 
to be sensitive to the relative direction of the mode propagation and the flow 
shear, specifically via the ratio $k/\(k_\parallel l_v\)$. 
A clear discussion of the dynamics of this parallel flow shear-driven mode was 
given by Catto \etal~\cite{catto}, among others. 
Perturbing the density causes a perturbed electric field, via $\delta n_e = \( e\phi / T \) n_0$. 
The $E \times B$ motion in the perturbed electric field, with direction 
defined by $k$, convects the background parallel flow. 
The perturbed flow caused by this convection enhances (for $k/k_\parallel l_v>0$) 
or opposes (for $k/k_\parallel l_v<0$) the parallel motion causing the density 
perturbation. Instability results when the rate of increase of the flow by the 
convection overcomes the deceleration due to the parallel pressure 
gradient --- this happens when $\omega_v^* > k_\parallel$.

The parallel and perpendicular flow shear are related, as noted in 
\sref{secresponse}, so the stability criterion \eref{instcondition} may be 
translated into a requirement on the Mach number. We define:
\be
\omega_v^* = \alpha M, \qquad
\alpha = \frac{3}{8}k{\bf \frac{\hat{e}_v\cdot \hat{z}}{\hat{e}_v\cdot \hat{y}}},
\label{defnalpha}
\ee
and the angles $\theta$ and $\theta_v$, giving the direction of mode propagation 
and of the background flow, with respect to the magnetic field:
\be
\tan \theta = \frac{k}{k_\parallel}, \qquad
\tan \theta_v = {\bf \frac{\hat{e}_v\cdot \hat{y}}{\hat{e}_v\cdot \hat{z}}}.
\ee
Thus \eref{instcondition} implies that, for instability:
\be
\alpha M > k_\parallel.
\ee
In the case when both $\omega_v^*$ and $M$ are positive, we obtain the instructive rearrangement:
\be
M > \frac{k_\parallel}{\alpha} = \frac{8}{3}\frac{\tan \theta_v}{\tan \theta},
\ee
(Note that we may always take $M>0$ without loss of generality due to the symmetry of the system.)
This form will prove convenient in \sref{secconvection} and simply indicates 
that for a background flow velocity with a finite parallel component, modes 
with a finite parallel wavevector require a stronger drive to become unstable. 
Note that the mode frequency is given by \eref{wvmodedr}.

\begin{figure}
\centering
\includegraphics[width=.5\textwidth]{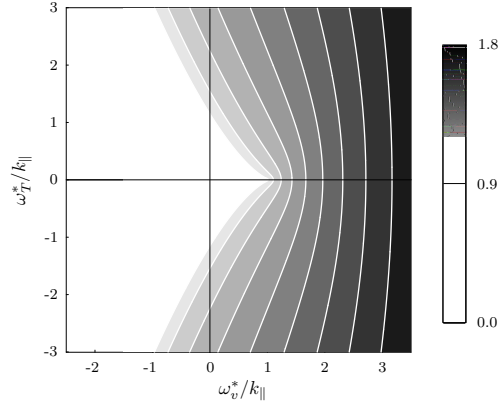}
\caption{Interaction of the ITG and PVG modes in the local dissipationless limit. Contours of the growth rate (divided by $k_\parallel$) of the fastest growing solution to \eref{wvwtboundary} are shown, with $\omega_n^* =0$.}
\label{figure2}
\end{figure}

Various stability boundaries may be derived from \eref{localdr}. 
It is illuminating to consider the two-dimensional stability boundary in 
the $\(\omega^*_v, \omega^*_T\)$-plane, which shows the impact of parallel 
velocity shear on the ITG instability. 
The dispersion relation in this case reduces to:
\be
\omega'^3  - \omega' k_\parallel \left(k_\parallel - \omega_v^*\right) + 
\frac{3}{8} k_\parallel^2 \omega_T^* =0.
\label{wvwtboundary}
\ee
Instability exists if either:
\be
\frac{\omega_v^*}{k_\parallel} > 1
\ee
or
\be
\lt |\frac{\omega_T^*}{k_\parallel}\rt | > \frac{16}{9\sqrt{3}}
\(1-\frac{\omega_v^*}{k_\parallel}\)^{3/2}.
\ee
This can be seen to reduce to the ITG stability criterion for 
$\omega_v^* =0$. The stability boundary is shown in \fref{figure2} and the asymmetric effect 
of background parallel flow shear on stability may be clearly seen.

\section{Parallel Flow Instability with $M=0$}
\label{secpardissip}

The dissipative terms which were neglected in the previous section will act to 
damp the perturbations at large distances along the field line. 
Analytical solutions demonstrating this may be obtained by considering the case 
with no perpendicular flow shear, $M=0$. 
For tractability we retain only viscous dissipation in the system. 
(As may be seen from \eqs{neq}{teq}, this corresponds to a system with an 
artificially large Prandtl number, $\nu_\perp/\chi_\perp \gg 1$.) 
The linearized equations then reduce to:
\bea
\frac{\p n}{\p t} + \frac{\p V}{\p z} = i\omega_n^* n,
\label{eqndiss}\\
\frac{\p V}{\p t} + \frac{3}{8}\frac{\p}{\p z}\(2 n+ T\) = i\omega_v^*n - 
\nu_k\(1+z^2\)V,\\
\frac{\p T}{\p t} + \frac{2}{3}\frac{\p V}{\p z} = i\omega_T^*n.
\label{eqtdiss}
\eea
No explicit time dependence appears, so we may look for a solution which varies 
as $\exp \(\gamma t\)$. 
The set can then be formed into a single mode equation, written here for the 
perturbed parallel velocity:
\be
\(\frac{\p}{\p{z}} - i \frac{\omega^*_v}{b}\) \frac{\p}{\p{z}} V - 
\frac{a}{b}\(\gamma + \nu_k + \nu_k z^2\) V = 0.
\label{eigen1}
\ee
The effect of the drives $\omega^*_T$ and $\omega^*_n$ are given by the 
parameters $a$ and $b$, defined as:
\bea
a = \gamma - i \omega_n^*, \\
b = 1 + \frac{i}{4\gamma}\(\frac{3}{2}\omega_T^* - \omega^*_n\).
\eea

It is convenient to define:
\be
V = s \( z \) \exp\(i \omega^*_v z / 2 b \) \exp\(\gamma t\),
\label{eqrapidosc}
\ee
Note that $s\( z \)$ must converge sufficiently rapidly as 
$|z|\rightarrow \infty$ to ensure that $V\rightarrow 0$ as 
$|z|\rightarrow \infty$. \Eref{eigen1} reduces to the form:
\be
\frac{\p{^2s}}{\p{z^2}} + \left[\frac{\omega_v^{*2}}{4b^2} - 
\frac{a}{b}\(\gamma + \nu_k\) - \frac{a}{b}\nu_k z^2\right] s = 0.
\ee
Bound state solutions (those satisfying the boundary conditions 
$V\rightarrow 0$ as $|z|\rightarrow \infty$) require that the real 
part of $\sqrt{a/b}$ be positive. Upon defining the variable:
\be
\xi = \(\frac{a \nu_k}{b}\)^{1/4} z,
\ee
the mode equation may be cast into quantum harmonic oscillator form:
\be
\frac{\p ^2 s}{\p \xi^2} + \left[ \epsilon - \xi^2 \right] s = 0.
\label{qhoform}
\ee
The parallel structure of the mode, described by $s\(z\)$, therefore has the form:
\be
s\(z\) = H_n\left[\(\frac{a \nu_k}{b}\)^{1/4} z\right]
e^{-\sqrt{a \nu_k/b}~(z^2/2)},
\ee
where $H_n$ are the Hermite polynomials. 
The mode thus corresponds to waves, of $n$ nodes in the parallel direction, 
trapped between dissipative regions far along the field line. 
There is also the overlying rapid oscillation, given by \eref{eqrapidosc}, so:
\be
V = H_n\left[\(\frac{a \nu_k}{b}\)^{1/4} z\right]
\exp\(-\sqrt{\frac{a \nu_k}{b}}~\frac{z^2}{2} + i \frac{\omega^*_v z }{ 2 b} \) \exp\(\gamma t\).
\label{fullvsoln}
\ee

The quantized eigenvalues in \eref{qhoform}, $\epsilon = 2n +1$, for positive integer $n \geq 0$, 
correspond to the allowed growth rates, such that:
\be
\frac{a}{b}\(\gamma + \nu_k\) + \(2n +1\)\sqrt{\frac{a \nu_k}{b}} - 
\frac{\omega_v^{*2}}{4 b^2} = 0.
\label{eigengrowth}
\ee
In the limit $\omega_T^* = \omega_n^* =0$, the effect of dissipation on the 
parallel flow shear driven mode is seen to produce a growth rate satisfying:
\be
\gamma^2 + \nu_k \gamma + \(2n+1\)\sqrt{\nu_k \gamma} = \frac{\omega^{*2}_v}{4}.
\label{wveigengrowth}
\ee
Thus a growing mode exists for all finite values of $\omega_v^*$ and the growth 
rate is increased for a given drive strength if the dissipation, $\nu_k$, is 
reduced. When $\nu_k =0$, we recover the relation $\gamma = \omega_v^*/2$ of 
\sref{secldr}. 
At large $k$ ($\nu_k \gg \omega^{*}_v$) the mode is still growing but dominated 
by viscosity $\gamma \sim \omega_v^{*2}/4\nu_k$.  
In reality large $k$ modes will be damped either by particle transport or 
finite Larmor radius corrections --- 
effects excluded from this analysis by our ordering.

\begin{figure}
\centering
\subfloat (a)
{\includegraphics[angle=-90, width=0.45\textwidth]{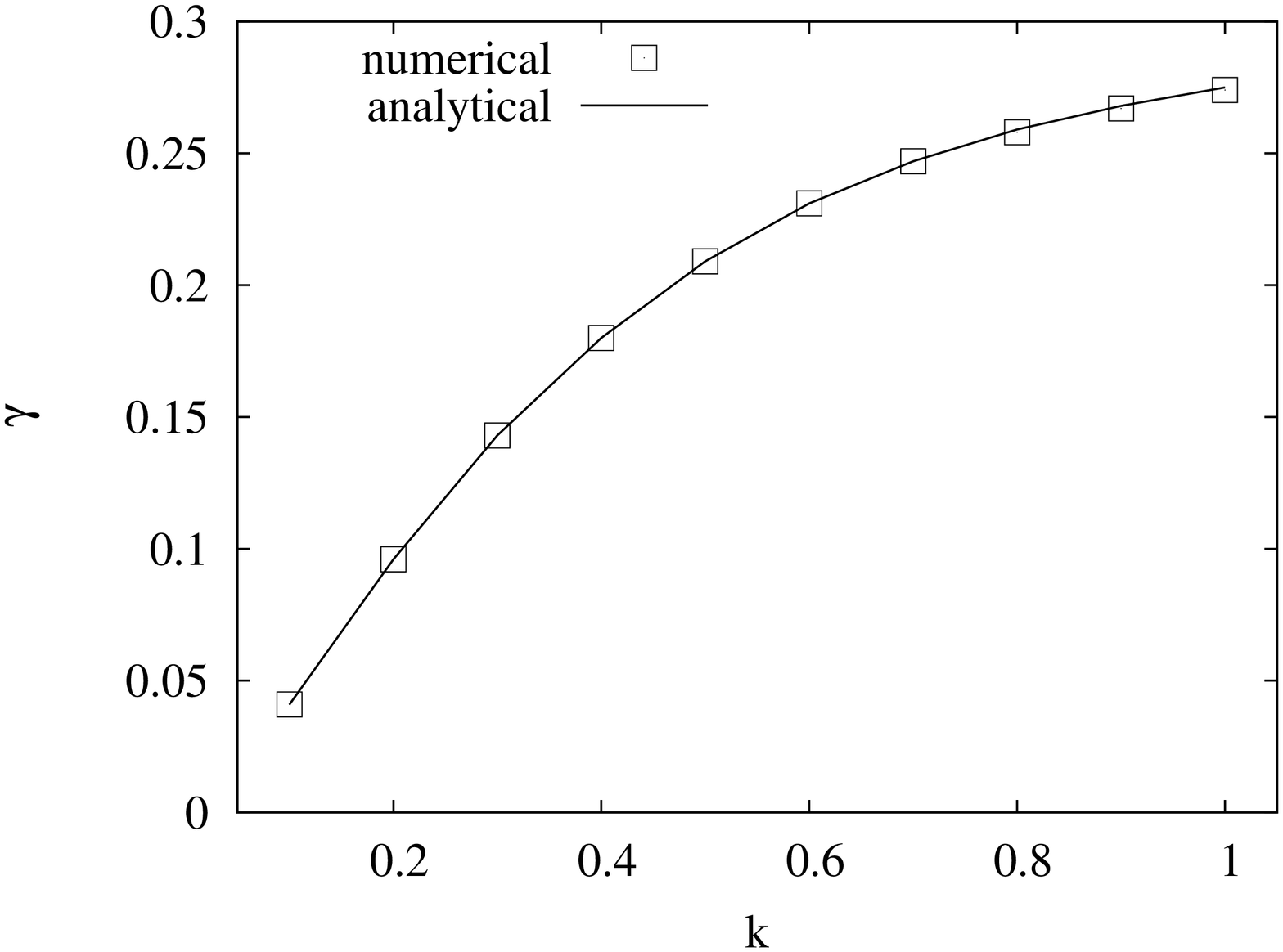}}
\subfloat (b){
\includegraphics[angle=-90, width=0.45\textwidth]{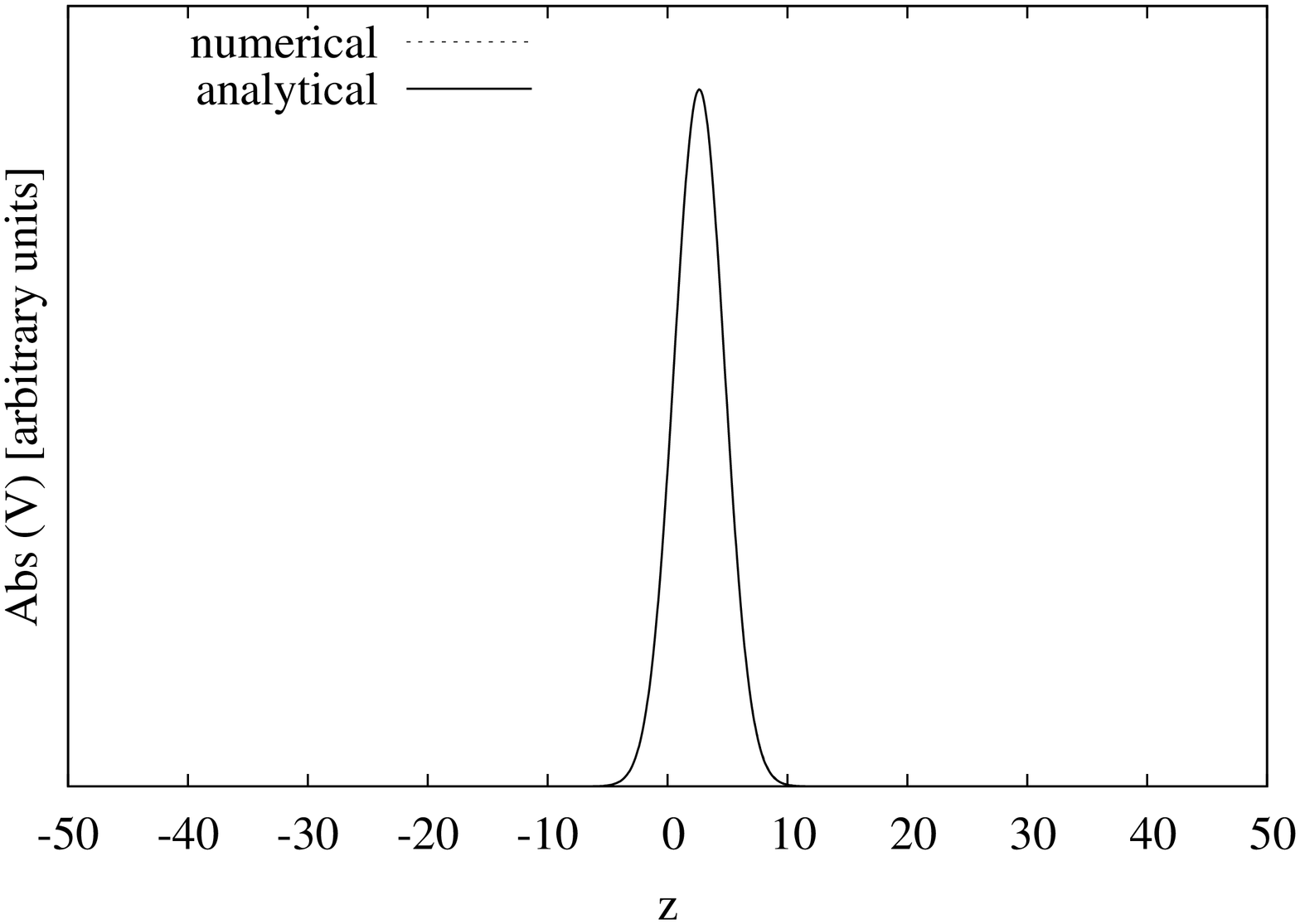}}
\caption{(a) Growth rate as a function of 
$k$. Squares are obtained from the numerical solution 
of the system of \eqs{neq}{teq}, with parameters: $l_s/l_n = 3,~l_s/l_T=6$, 
$l_s/l_v = 9$, $M=0$, $\nu=3$ and $\chi=0$. The corresponding analytical solution~\eref{eigengrowth} for $n=0$ 
is also shown (solid line). (b) The numerically obtained velocity eigenfunction (absolute value) for 
$k=0.3$ (dotted line)
with the analytical expression~\eref{fullvsoln} (solid line).}
\label{figure3}
\end{figure}

Shown in \fref{figure3} is a validation of equations
\eref{eigengrowth} and \eref{eqrapidosc} 
against a numerical solution of the full set 
of \eqs{neq}{teq} in the limit $M=0,~\chi=0$. 
(A description of the code is given in \sref{secnumerics}, where results 
of numerical investigations of the full linearized system defined 
in~\sref{seclinsys} are presented.) The numerical solution displayed is the fastest growing mode found, whilst the analytical solution plotted 
corresponds to the case $n=0$. 

\section{Impact of Convection}
\label{secconvection}

As was seen in \sref{secresponse}, the perpendicular flow shear acts to convect 
perturbations along the system. 
Dissipation, aided by the strong magnetic field shear, acts to remove all but 
the density perturbation at large distances along the field lines. 
Thus we may anticipate that rapid convection will sweep growing instabilities 
into the dissipative region and force their decay. 
This is motivated in the following section by considering the characteristics 
of the linear system, \eqs{neq}{teq}. 
In such cases, the system would be considered linearly stable, as no 
exponentially growing eigenmodes would be found. 
However, an unstable mode would be able to grow for a finite time. 
The form of such transitory solutions, which begin to grow exponentially 
but are then forced to decay, is determined in \sref{sectransitory}. 
As discussed in the Introduction, it is of interest to determine if transitory 
solutions grow to sufficient amplitude to trigger non-linear effects.


\subsection{Characteristic Equations}
\label{seccharacteristics}

The linear set of equations describing the system, \eqs{neq}{teq}, can be 
conveniently written in characteristics form. Defining:
\be
S=\frac{3}{2}T-n, \qquad C^\pm = V \pm \frac{3}{4}\(n+\frac{T}{2}\),
\ee
we obtain:
\bea
\label{eqnentrop}
\fl \[\frac{\p}{\p t} + M\frac{\p}{\p z}\]S = i\frac{\omega_S^*}{2}
\(C^+ - C^- -\frac{S}{2}\) - \frac{1}{2}\chi_k \(1+z^2\)
\(C^+ - C^- +\frac{3S}{2}\),\\
\label{eqncsound}
\fl \[\frac{\p }{\p t} + \(M \pm 1\)\frac{\p }{\p z}\] C^\pm =
i\frac{\omega_\pm^*}{2}\(C^+ - C^- -\frac{S}{2}\) \nonumber \\
- \frac{1}{2} \nu_k \(1+z^2\) \( C^+ + C^-\) \mp \frac{1}{8} 
\chi_k \(1+z^2\) \(C^+ - C^- +\frac{3S}{2}\),
\eea
where
\bea
\omega_S^* = \[\frac{3}{2}\omega_T^* - \omega_n^*\], \nonumber \\
\omega_\pm^* = \[\omega_v^* \pm \frac{3}{4}\(\frac{\omega_T^*}{2} + \omega_n^*\)\].
\eea
These characteristics represent three waves: the entropy mode propagating the 
perturbed specific entropy, $S$, at speed $M$, the forward propagating sound
 wave propagating $C^+$ at speed $M+1$ and the backward propagating sound wave 
propagating $C^-$ at speed $M-1$. 
The terms on the right hand sides of \eref{eqncsound} couple the three 
waves but do not change their propagation speeds. 
An initial perturbation which is localized in $z$ between $z=a$ and $z=b$ at 
time $t=0$ ({\it i.e.} the function describing the initial perturbation has 
compact support in $z$) must be localized between $z=a + (M-1)t_1$ and 
$z=b + (M+1)t_1$ at time $t=t_1$.  
Eigenmodes can form when $M<1$, by the combination of oppositely travelling 
waves coupled by the right hand sides of \eref{eqncsound}.

As $M$ is increased, we see from \eref{eqncsound} that the speed of the forward 
travelling characteristic is enhanced, whilst that of the backward is reduced. 
Thus, when $M>1$ is reached, all of the characteristics of the system are 
forward going, due to the convective effect of the perpendicular shear of the 
background flow. 
In this case no eigenmode may be formed in the system. 
Indeed the initial perturbation is swept forward, since $z=a + (M-1)t_1$ 
and $z=b + (M+1)t_1$ both increase with the time $t_1$. 
This discussion is borne out by the numerical solutions presented in 
\sref{secnumerics}, see \fref{figure6}.
As described in \sref{secpardissip}, velocity and temperature perturbations will 
damp at large distances along the field line. 
Note that the density perturbation will remain, due to the form of \eref{neq}, 
as will also be seen in \sref{secnumerics}.
Therefore, for $M>1$, any initial unstable perturbation will be swept into the 
dissipative region and forced to decay. 
Whilst such a perturbation can grow exponentially for a finite time, 
its behaviour would not be captured by a traditional eigenvalue analysis. 
In the following section we consider the evolution of such a transitory 
perturbation.


\subsection{Transitory Solution}
\label{sectransitory}

In order to determine the behaviour of transitory instabilities, we consider a 
system with strong shear of the perpendicular background flow, that is $M \gg 1$. 
An analytic solution may be obtained upon retaining only the driving term due to 
the parallel flow shear, which is associated with the strong convection via 
\eref{defnalpha}, and formally neglecting the effect of thermal conductivity, 
as in \sref{secpardissip}. 
The solution in this limit captures the nature of the transitory behaviour, 
which arises due to the strong convection rather than being dependent on the 
details of the dissipation mechanism. 
Therefore we consider in this section the transitory evolution of the 
$\omega_v^*$-driven instability described in \sref{secldr}. 
The transitory behaviour of the full system was investigated numerically 
and the results are presented in \sref{secnumerics}.

It is convenient to analyse the system in the moving frame, 
using the coordinates $\(z_0,t\)$, where $z_0 = z - Mt$. 
With $\omega_T^* = \omega_n^* = \chi_k = 0$, the system of \eqs{neq}{teq} 
may be reduced to a single differential equation, written here in terms of 
the parallel velocity perturbation:
\be
\frac{\p^2 V}{\p t^2}  = \frac{\p^2 V}{\p{z_0^2 }} - i \omega_v^* 
\frac{\p V}{\p z_0} - \nu_k \frac{\p}{\p t}\left[\(1 + z^2\) V\right].
\label{transmodeeq}
\ee
The time derivative is to be taken at constant $z_0$. 
The rapid parallel oscillation inherent in this instability, which was present
 when $M=0$ (see \sref{secpardissip}), may be removed upon substituting: 
$ V\(z_0,t\) = \tilde{\cal{V}}\(z_0,t\) \exp\left[i \omega_v^* z_0/2 \right]$. 
This reduces \eref{transmodeeq} to:
\be
\fl \frac{\p^2 \tilde{\cal{V}}}{\p t^2}  = 
\frac{\p^2 \tilde{\cal{V}}}{\p{z_0^2 }} + \frac{\omega_v^{*2}}{4}\tilde{\cal{V}} 
- \nu_k \left[\(1 + z_0^2 + 2 z_0 M t + M^2t^2\) \frac{\p \tilde{\cal{V}}}{\p{t}} 
+ 2 M \(z_0 + Mt\) \tilde{\cal{V}}\right].
\label{vtildemodeeq}
\ee

We anticipate that the initial exponential growth of the instability in time, 
associated here with the explicit $\omega_v^*$ term as seen in \sref{secldr}, 
will be overcome after a finite time by dissipative effects when $M \gg 1$. 
This exponential behaviour may be conveniently characterized by introducing the 
time dependent ``growth rate" $\gamma\(t\)$, such that: 
$\tilde{\cal{V}}\(z_0,t\) = {\cal{V}}\(z_0,t\) 
\exp \left[\int^t\gamma\(t'\)dt'\right]$, satisfying:
\bea
\fl \frac{\p^2{\cal{V}}}{\p t^2} + 2 \gamma \frac{\p {\cal{V}}}{\p t} + 
\gamma^2 {\cal{V}} = \frac{\p^2 {\cal{V}}}{\p{z_0^2 }} + \nonumber \\
\fl \quad\quad\frac{\omega_v^{*2}}{4}{\cal{V}}  - 
\nu_k \left[\(1 + z_0^2 + 2 z_0 M t + M^2t^2\) \(\gamma {\cal{V}} + 
\frac{\p {\cal{V}}}{\p{t}}\) + 2 M \(z_0 + Mt\) {\cal{V}}\right].
\label{vmodeeq}
\eea
This evolution equation may be expanded in the large parameter, $M$. We seek a solution where:
\bea
\frac{\p {\cal{V}}}{\p{z_0 }} \sim \mathcal O\(1\){\cal{V}} \qquad {\rm and} 
\qquad \frac{\p {\cal{V}}}{\p{t}} \sim \mathcal O\(M^{1/2}\){\cal{V}}. 
\label{transorder}
\eea
Remembering that for short times $\gamma\(t\)$ will approximate the local 
solution of \sref{secldr}, where $\gamma \sim \omega_v^* \propto M$ , 
the leading terms of \eref{vmodeeq} are of $\mathcal O \(M^2\)$ and yield:
\be
\gamma^2  = \frac{\omega_v^{*2}}{4} - \nu_k M^2 t^2 \gamma.
\ee
Hence the function $\gamma$ is given by:
\be
\frac{\gamma}{\gamma_0} = -\tau^2 + \sqrt{\tau^4 + 1}, \qquad \tau = \sqrt{\frac{\nu_k}{2\gamma_0}}Mt,
\label{gammadefn}
\ee
where we have taken the positive root of the radical, as the behaviour of the 
fastest growing instability is of most interest, and we have normalized to the 
local value of the growth rate in the case of instability, 
$\gamma_0 = \omega_v^* / 2$ (see \sref{secldr}). 
For long times, $\gamma / \gamma_0 \rightarrow (2 \tau^2)^{-1}$ and exponential 
growth will no longer dominate the evolution of the perturbation. 
Therefore we may identify that the instability will grow exponentially for a 
characteristic time $t_c = \sqrt{\gamma_0}/\(\sqrt{\nu_k} M \) 
\propto 1/\sqrt{M}$, during which time the initial perturbation will be 
amplified by a factor of order $\exp\left[\gamma_0 t_c\right] \sim 
\exp\left[a\sqrt{M}\right]$, where $a$ is independent of $M$. 
Clearly at large velocity shear ($M$) the transitory perturbations can 
grow significantly.

The amplitude of the perturbation will turnover around $t_c$ and begin to decay 
due to dissipation. 
Thus the function ${\cal{V}}$ will change on the timescale $\tau_c$ 
(slower than the growth time $\gamma_0^{-1}$) as anticipated in 
\eref{transorder}. 
The behaviour of ${\cal{V}}$ will be governed by the next significant terms 
in \eref{vmodeeq}, which are of $\mathcal O \(M^{3/2}\)$:
\be
2 \gamma \frac{\p {\cal{V}}}{\p t} = - \nu_k Mt 
\left[ Mt\frac{\p {\cal{V}}}{\p t} + 2z_0 \gamma {\cal{V}} + 2M {\cal{V}}\right].
\label{sevolution}
\ee
This is simply a first order equation for the time dependence of ${\cal{V}}$, in which $z_0$ is a parameter, therefore with $\gamma$ and $\tau$ given by \eref{gammadefn}:
\bea
\fl {\cal{V}}\(z_0,t\) & = \overline{{\cal{V}}}\(z_0,t\) 
\exp\left[-\int\frac{\(1 + \frac{z_0 \gamma}{M}\)}{\(\frac{\gamma}{\gamma_0} + 
\tau^2\)}d\(\tau^2\) \right] \nonumber \\
\fl & = \overline{{\cal{V}}}\(z_0,t\)\(\tau^2 - \sqrt{1+\tau^4}\)\exp\left[-\frac{z_0 \gamma_0}{M}\(\tau^2 - \sqrt{1+\tau^4}\)\right],
\eea
where we have used the substitution $\tau^2 = \sinh \vartheta$ and noted that $\sinh \vartheta - \cosh \vartheta = e^{-\vartheta}$ to obtain the final form. 
The function $\overline{{\cal{V}}}$ has been introduced to capture the remaining 
finite spatial and temporal dependence of the initial perturbation amplitude, 
present in \eref{vmodeeq}.

The perturbed parallel velocity for $M \gg 1$ thus takes the final form: 
\be
\hskip -0.5in V\(z_0,t\) = \overline{{\cal{V}}}\(z_0,t\)\(\tau^2 - 
\sqrt{1+\tau^4}\)\exp\left[i\frac{\omega_v^*}{2}z_0 + 
\frac{z_0}{M}\gamma\(t\) + \int \gamma \(t\) dt \right].
\ee
In the initial stage of the instability, as $M\gg1$ the final term in the 
exponential will dominate the evolution, giving exponential growth of the 
perturbation amplitude with $\gamma\(t\) \approx \gamma_0$. 
However, at times longer than the saturation time: $t\gg t_c 
\sim 1 / \sqrt{M}$, $\gamma\(t\)$ varies as $t^{-2}$ 
and the asymptotic form of $V\(z_0,t\)$ becomes:
\bea
\fl V\(z_0,t\) = -\overline{{\cal{V}}}\(z_0,t\)\frac{\gamma_0}{\nu_k M^2}\(\frac{1}{t^2}\) \exp{(\Phi)} \nonumber \\
\fl \Phi = \left[i\frac{\omega_v^*}{2}z_0 + \frac{z_0}{M}\frac{\gamma_0^2}{\nu_k M^2}\(\frac{1}{t^2}\) + \frac{\gamma_0^{3/2}}{\sqrt{\nu_k/2} ~ M}\(1.236 - \frac{\gamma_0^{1/2}}{\sqrt{2\nu_k} ~ M t}\)\right].
\label{asymptrans}
\eea
The exponential factor therefore saturates and the amplitude evolution in the decaying phase is determined by the algebraic prefactor, which decays as $t^{-2}$. Thus, not only can transitory perturbations reach significant amplitude during their initial exponential growth, they linger in the system, decaying in amplitude only algebraically with time. The density perturbations do not decay after an initial amplification and only effects outside this analysis (either particle transport or finite Larmor radius effects) cause their eventual decay.  Note that the total number of e-foldings is: $~0.1 \sqrt{k M/\nu_\perp}\(\cot \theta_v \)^{3/2}$.
As expected, large parallel flow shear gives large transitory growth.


\section{Numerical Results}
\label{secnumerics}

The results of numerical investigations of the linearized system defined in 
\sref{seclinsys} will now be presented. 
We integrate \eqs{eqnentrop}{eqncsound} numerically with a second-order accurate 
upwind scheme, 
which naturally captures the forward or backward propagation of the characteristics.
The equations are solved in a box of size $-z_{inf}<z<z_{inf}$, 
where $z_{inf}$ varies between $50$ and $200$ depending on the parameters.
The typical resolution is $\Delta z=0.1$. 
Convergence tests were performed to establish that the size of the domain and the 
resolution employed were appropriate; a benchmark of the code against an analytical 
solution has been shown in~\fref{figure3}.
The perpendicular viscosity~\eref{nu_chi_perp} is fixed at $\nu_{\perp} = 3.0$ for 
all cases presented: this relatively large value of $\nu_{\perp}$ ensures 
the consistency of our orderings.
The analytically defined instability drives (27) depend on the wavenumber $k$ 
as does the growth rate. 
Therefore, in this section we present the stability results as functions of the 
physical instability drives, the gradient scale length ratios 
$(l_s/l_n, l_s/l_T, l_s/l_v)$.
In figures showing the maximum growth rate for a given set of physical parameters, 
the growth rate has been maximised over the wavenumber $k$, using the well known 
Brent's method~\cite{Numerical_Recipes}. Finally, for clarity we remove the density gradient drive throughout this section, that is $l_s/l_n =0$ in all cases. The results of tests with finite values of the density gradient did not alter the conclusions presented here. 

\begin{figure}[h!]
\centering
\includegraphics[angle=-90, width=0.5\textwidth]{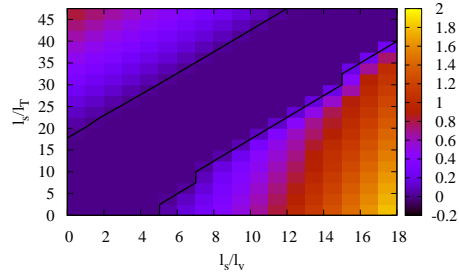}
\caption{Contour plot of the maximum growth rate as a function of temperature 
and parallel velocity shear drive strengths. $M=0.3$ 
and $l_s/l_n=0 $. Solid lines indicate zero growth rate.}
\label{figure4}
\end{figure}

In \fref{figure4} we illustrate the effect of the parallel velocity gradient on 
the ITG instability. 
A contour plot of the growth rate of the most unstable mode is shown as a 
function of the temperature, $l_s/l_T$, and parallel velocity shear, 
$l_s/l_v$, drives, at fixed convective velocity $M = 0.3$. 
Note that $l_s/ l_v = c_s M \cot \theta_v$, so the $x$-axis represents 
variation of both the angle and the magnitude of the flow. 
As we have taken $M < 1$, eigenmodes may form in the system and the growth rate 
is well defined --- see \sref{seccharacteristics}.
As seen, for a fixed value of $l_s/l_v$, the temperature gradient has to be 
raised above a certain critical threshold for the ITG to be unstable.
Conversely, we see that the minimum value of the velocity gradient required 
for instability shifts to the right on the plot as the temperature gradient 
is raised. This figure thus illustrates the complex interplay between the 
two different drives and 
highlights the fact that increasing one drive may have a stabilising effect on the 
system if the other drive is kept fixed.
Beyond $M = 1$ no eigenmodes are formed in the system 
and no growth rate may be determined in the usual way, as discussed in 
\sref{seccharacteristics}.

\begin{figure}[h!]
\centering
\subfloat (a){\includegraphics[angle=-90,width=0.45\textwidth]
{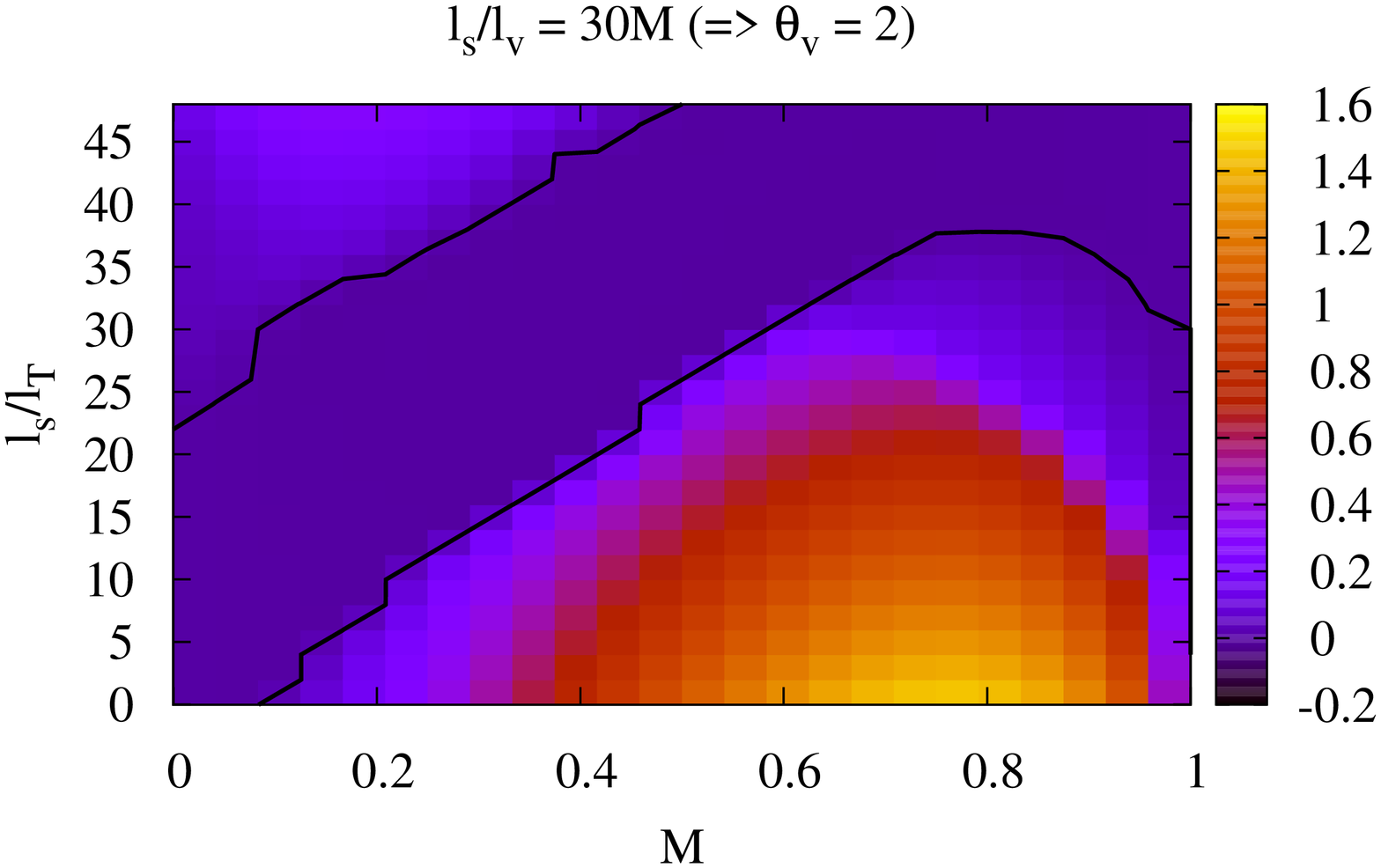}}
\subfloat (b){\includegraphics[angle=-90,width=0.45\textwidth]{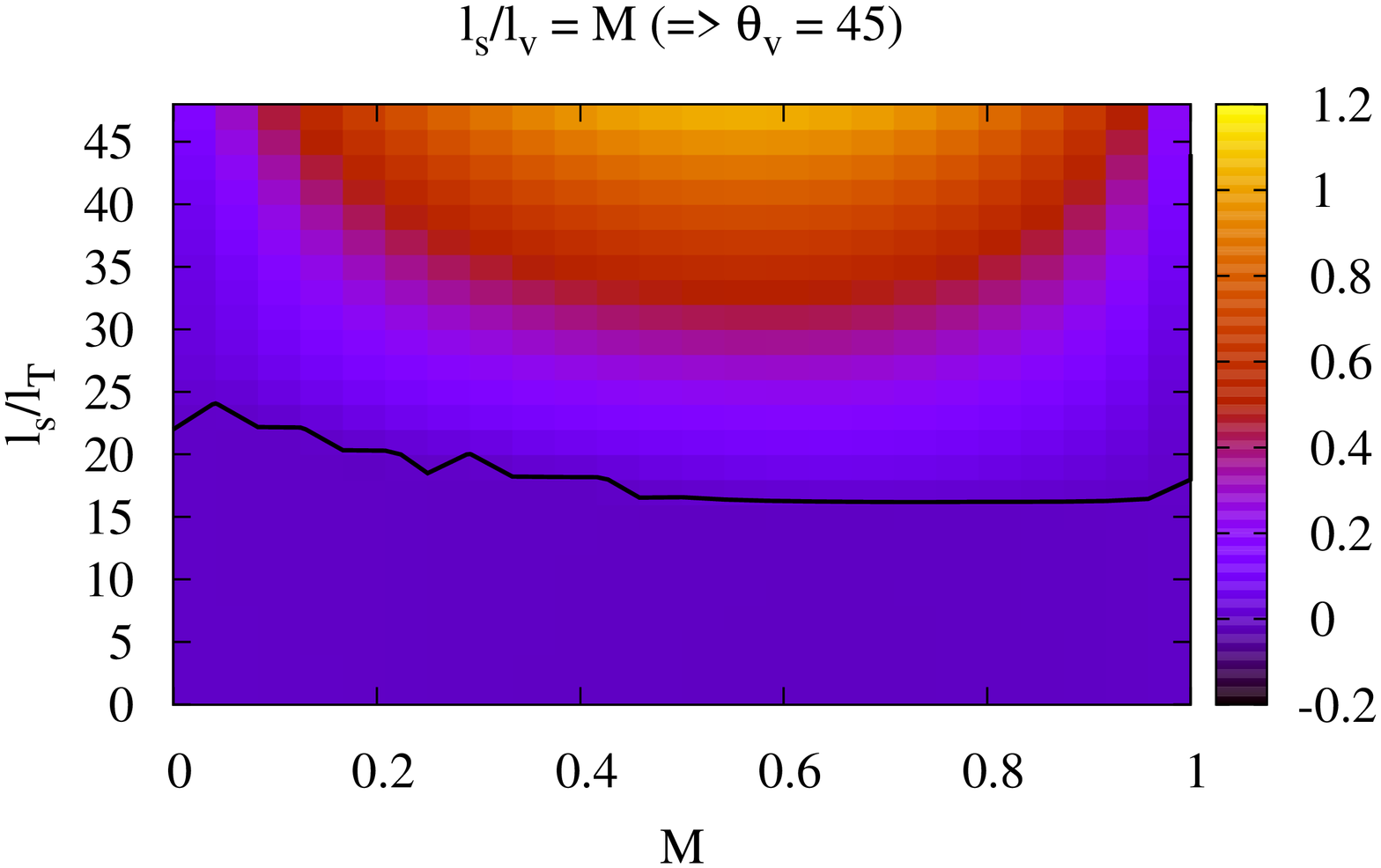}}
\caption{Contour plots of the maximum growth rate as a function of 
the temperature gradient drive and $M$ for 
$l_s/l_n=0$ and  
(a) $l_s/l_v=30M$, corresponding to $\theta_v=2\textdegree$; 
(b) $l_s/l_v=M$, corresponding to $\theta_v=45\textdegree$.
For $M>1$ no eigenvalues can be found. Solid lines indicate zero growth rate.}
\label{figure5}
\end{figure}

The more realistic case of fixed flow angle $\theta_v$ is exemplified 
in~\fref{figure5}, 
where we show contour plots of the maximum growth rate as a function of 
$l_s/l_T$ and $l_s/l_v$.
We depict two cases: \fref{figure5}(a) has $\theta_v=2\textdegree$, so
the flow is nearly 
parallel to the magnetic field, which is representative of a conventional 
large aspect ratio tokamak, whilst \fref{figure5}(b)
corresponds to the case $\theta_v=45\textdegree$, which is more 
pertinent to the case of a Spherical tokamak.
These plots illustrate particularly the stability threshold appearing at $M = 1$, 
which has been discussed in \sref{secconvection}.
They also show clearly the same non-trivial features seen
in~\fref{figure4}: we see that 
stable, or weakly unstable regions 
can also be found for $M<1$ in certain ranges of the drive strengths.
Consider, for example, fixed $M$ 
and thus fixed $l_s/l_v$. We see that as the ITG drive $l_s/l_T$ is 
increased, the maximum 
growth rate in the system (which corresponds to a pure PVG mode at $l_s/l_T=0$) 
actually decreases at first until a region of stability is reached;
this arises because the ITG and the 
PVG instabilities compete, rather than reinforce each other. 
As $l_s/l_T$ is increased even further, 
$\gamma_{max}$ eventually starts growing again and the system gradually 
transitions to a pure ITG mode.

Notice also that as $l_s/l_T$ increases, the $M$ value required to enter 
the stability band varies and a region of instability does not necessarily 
reappear as $M$ is increased further, depending on the angle of the flow. 
This is consistent with the result obtained by 
Waelbroeck~\etal~\cite{waelbroeck1}, who showed numerically that the value of 
perpendicular velocity shear (essentially the value $M$ used here) required for 
stability increased as $l_s/l_T$ increased, for $l_s/l_v$ in the region of $4M$. 
Note that the marginal stability curves presented in~\cite{waelbroeck1} are 
consistent with $M \lesssim 1$, upon recognising that the sound speed is 
defined there for cold ions and that the stability curves are determined for 
$T_e \neq T_i > 0$. 
As kinetic effects were retained, the precise definition of the sound speed 
used here for M is somewhat too simplified to reproduce the stability 
boundary of ~\cite{waelbroeck1} exactly. 

\begin{figure}
{\quad\qquad $M=0.3$ \qquad\qquad\qquad\qquad\qquad\quad $M=1.2$}\\
\centering
\subfloat{$n$}{
\vspace{-0.5cm}
\includegraphics[angle=-90,width=0.45\textwidth]{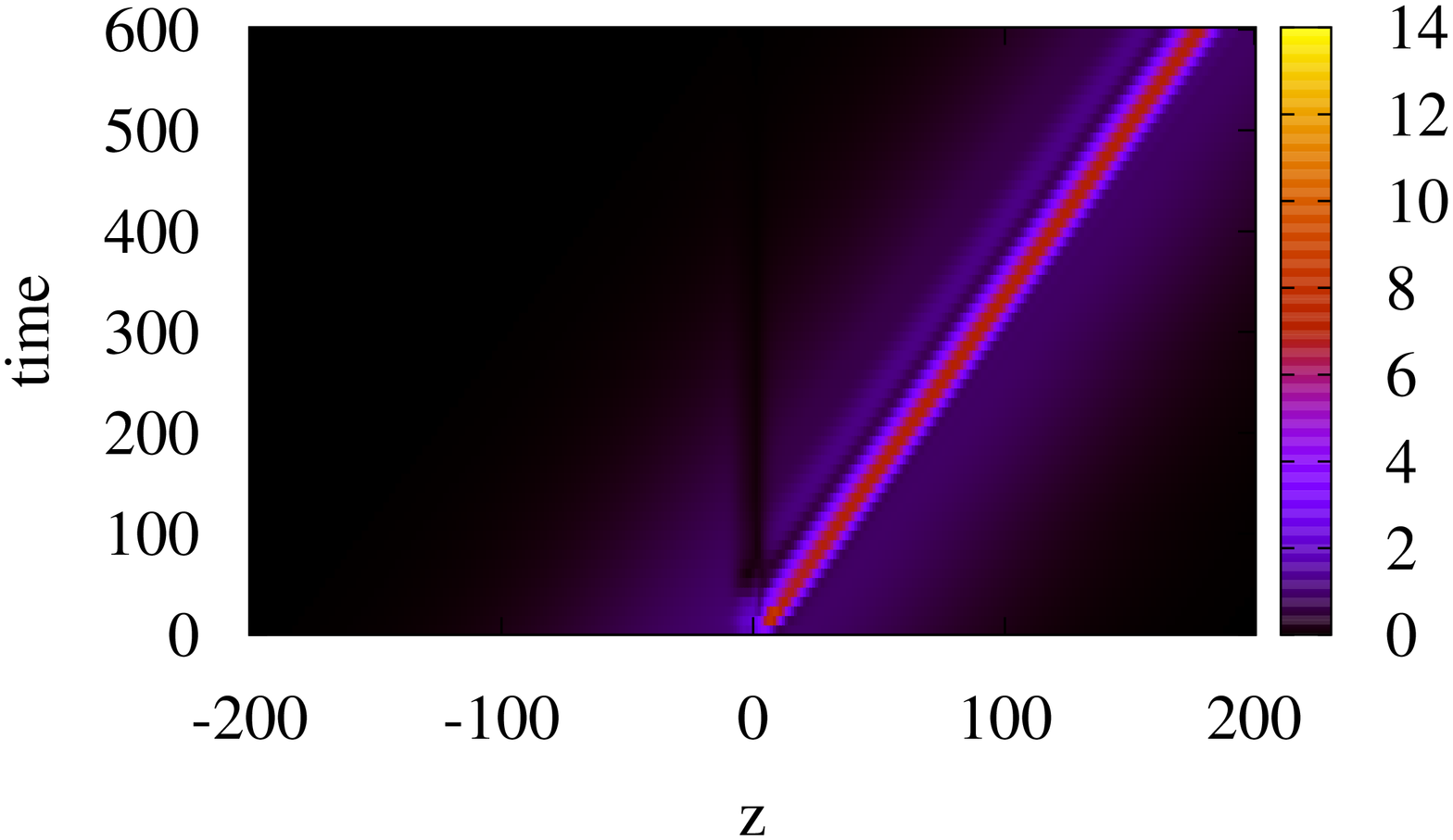}
\includegraphics[angle=-90,width=0.45\textwidth]{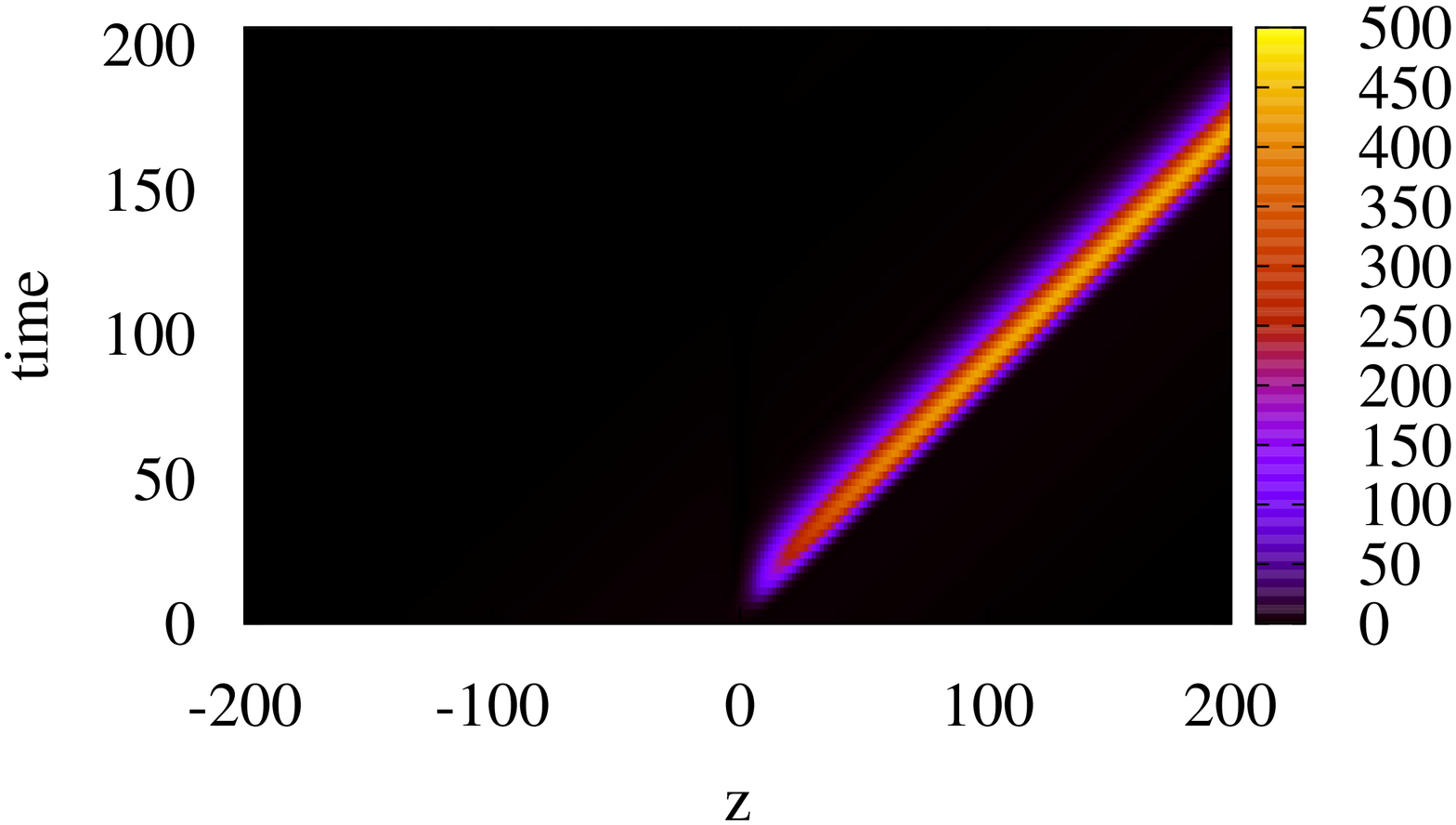}}\\
\subfloat{$T$}{
\vspace{-0.5cm}
\includegraphics[angle=-90,width=0.45\textwidth]{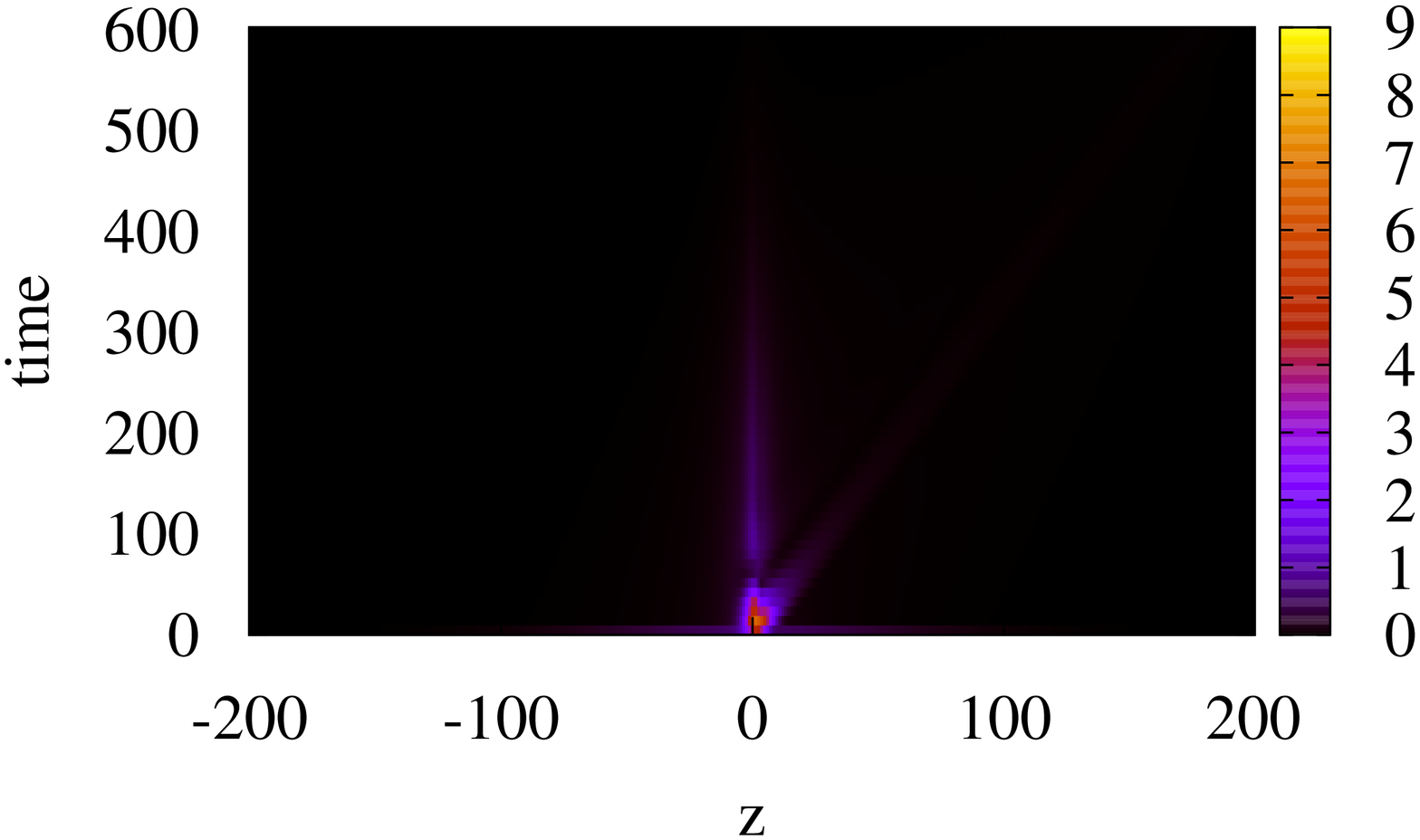}
\includegraphics[angle=-90,width=0.45\textwidth]{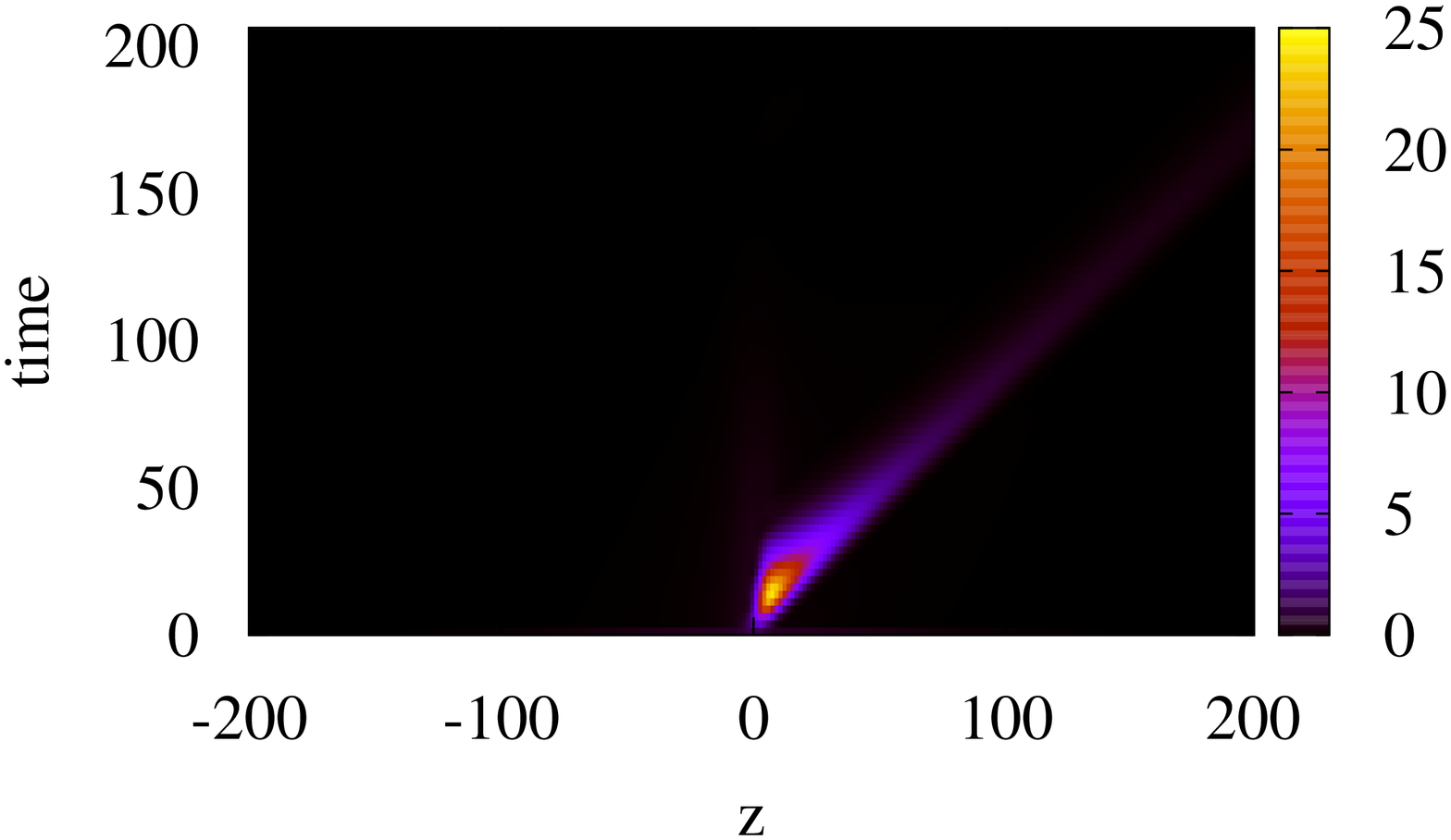}}\\
\subfloat{$V$}{
\vspace{-0.5cm}
\includegraphics[angle=-90,width=0.45\textwidth]{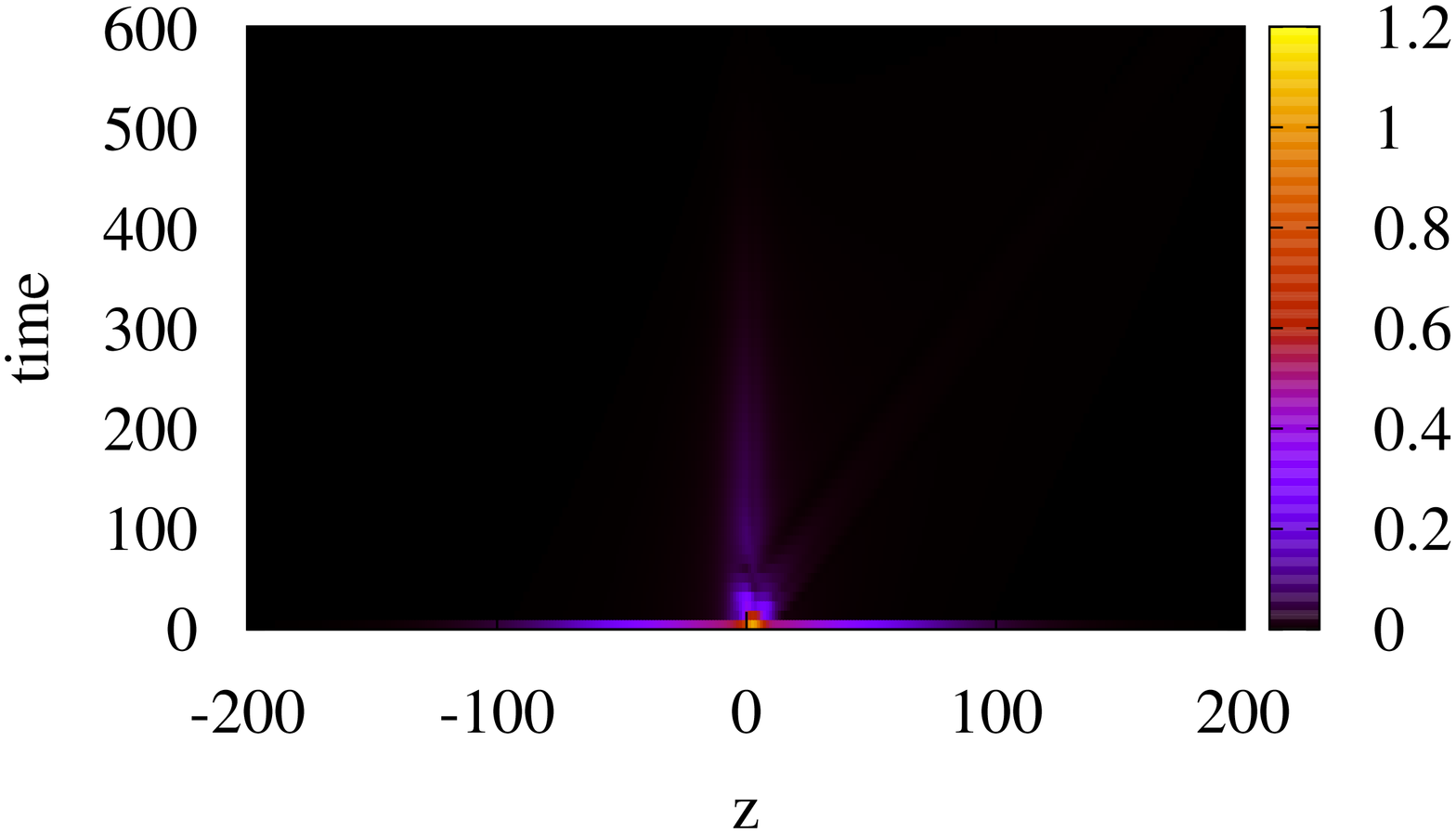}
\includegraphics[angle=-90,width=0.45\textwidth]{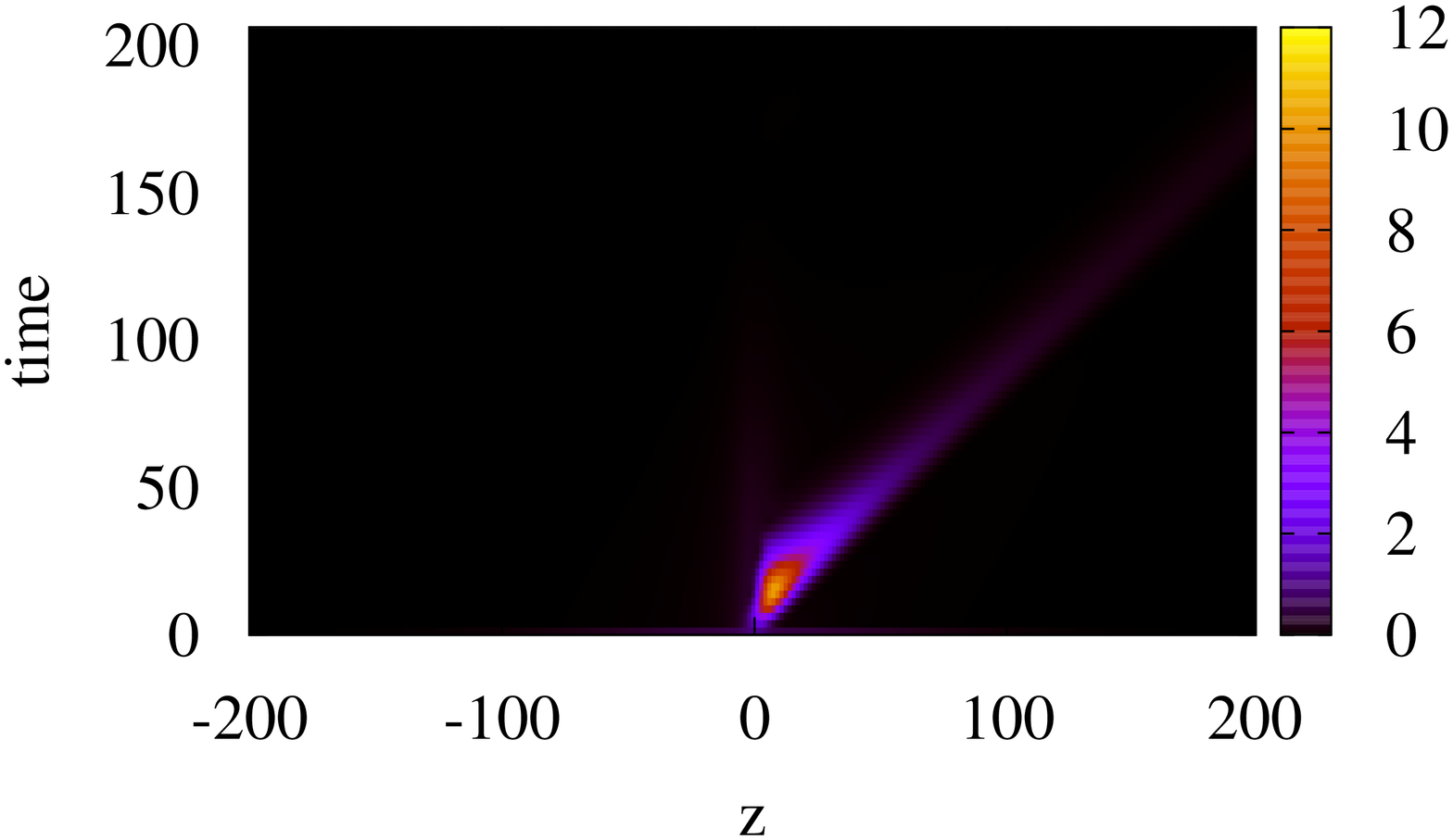}}
\caption{From top to bottom: time history plots of density, temperature and 
velocity (absolute values) for $l_s/l_n=0$, $l_s/l_T=30$,
$l_s/l_v=30M$ and $k=0.1$.
Left column shows case $M=0.3$ cf. \fref{figure5}(a);
right column shows case $M=1.2$. 
Each field is normalized to its maximum amplitude at $t=0$ so colors show 
the amplification factor of the initial perturbation.}
\label{figure6}
\end{figure}

\begin{figure}
\centering
\subfloat{$n$}{
\includegraphics[angle=-90,width=0.3\textwidth]{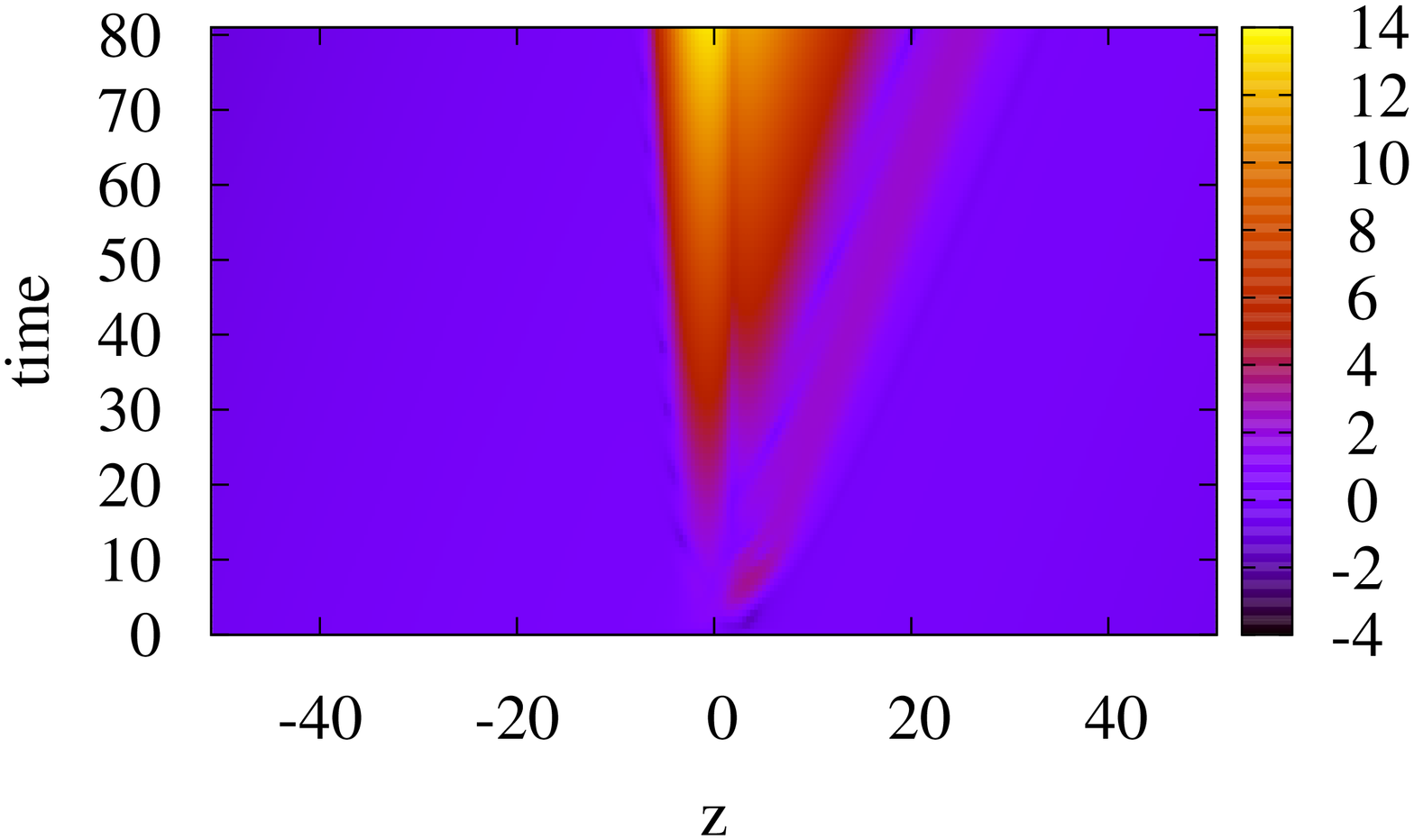}}
\subfloat{$T$}{
\includegraphics[angle=-90,width=0.3\textwidth]{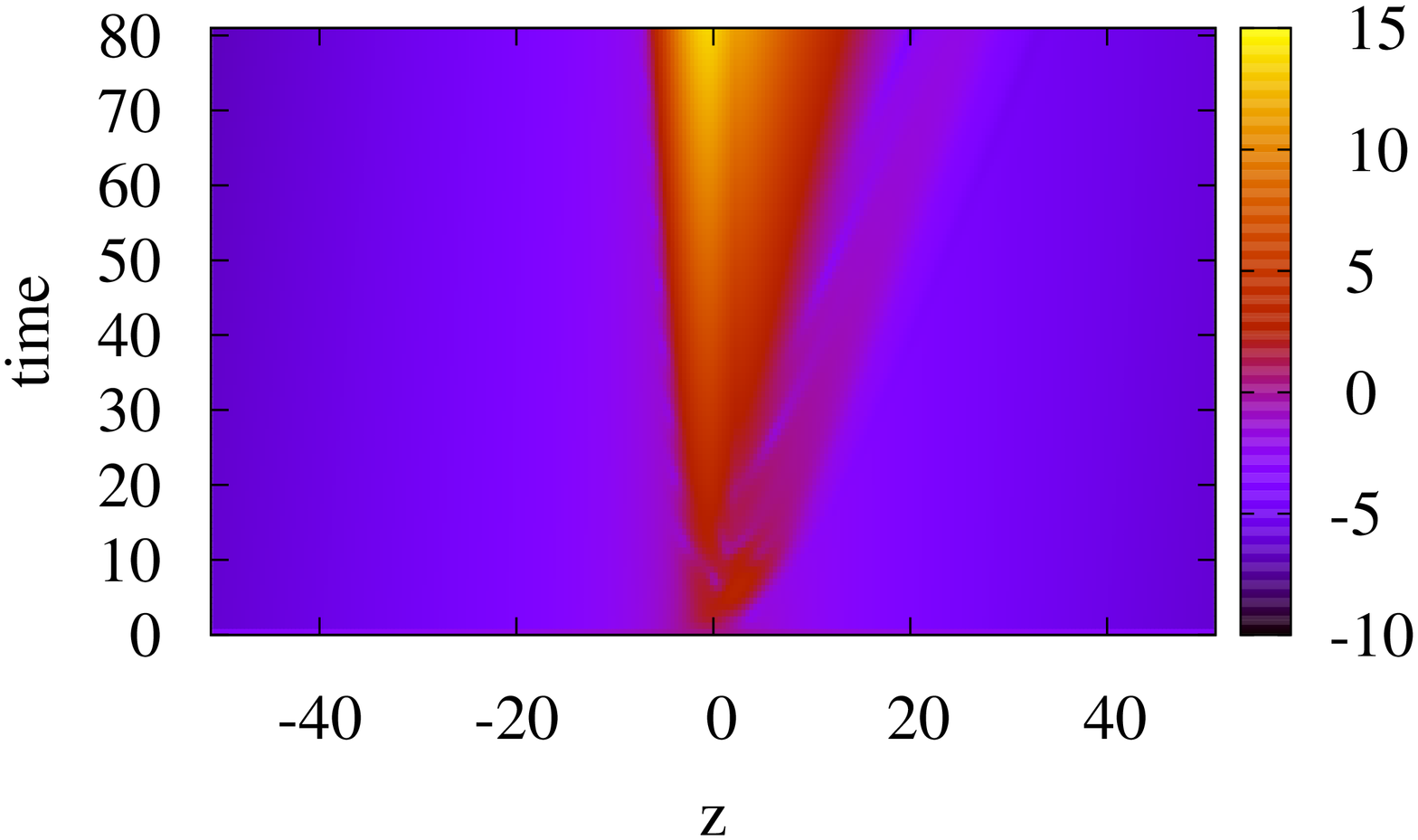}}
\subfloat{$V$}{
\includegraphics[angle=-90,width=0.3\textwidth]{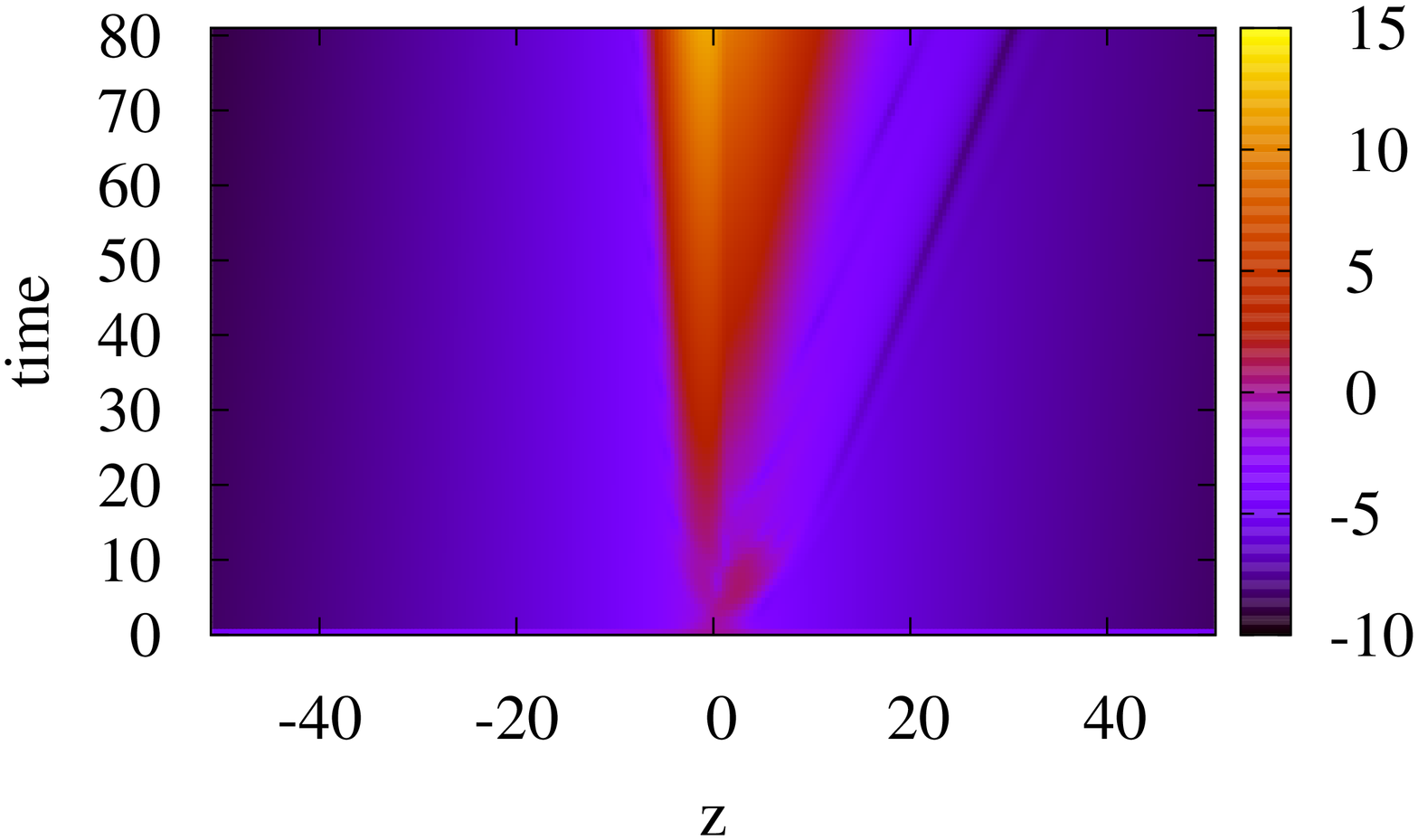}}
\caption{From left to right: time history plots of density, temperature and 
velocity (absolute values, log scale) for $l_s/l_n=0$, $l_s/l_T=45$ 
and $l_s/l_v=30M$ cf. \fref{figure5}(a). 
Here $k=k_{max}=0.3$, which is the fastest growing mode ($\gamma_{max}=0.17$).
Only the central fraction of the simulation box shown for clarity (simulation domain
is $-200<z<200$).
Each field is normalized to its maximum amplitude at $t=0$ so colors show 
the amplification factor of the initial perturbation.}
\label{figure7}
\end{figure}

The evolution of the perturbed fields in the
two distinct types of stability region either side of $M = 1$ is illustrated in~\fref{figure6}.
Taking $l_s /l_v = 30 M$, as in \fref{figure5}(a), and 
choosing $l_s/l_T = 30$ and $k = 0.1$, the left column of \fref{figure6} shows 
the case $M = 0.3$ which lies in the stability band of \fref{figure5}(a), 
whilst the right column has $M = 1.2$.
Even though both cases are formally stable, since there is no 
exponentially growing eigenmode, there is an amplification 
of the initial perturbation in both cases. This amplification is 
more significant in the case of $M=1.2$, particularly for the density.
Notice in particular that whereas the temperature and velocity  
fluctuations are eventually damped, the density perturbation is seen  
to remain in the system, as anticipated in \sref{seccharacteristics}, due to the lack of explicit dissipation in \eref{neq}.
For contrast, the perturbed fields in the case of an unstable eigenmode from 
\fref{figure5}(a), with $M = 0.3$ and $l_s/l_T = 45$, are shown in~\fref{figure7}. 
Here $k = 0.3$, corresponding to the fastest growing mode.


\section{Discussion and Conclusions}
\label{conclusions}

In this paper we have investigated the stability of ion temperature gradient 
and parallel velocity gradient driven modes in a sheared magnetic field with 
parallel and perpendicular sheared flows. 
The instability takes the form of twisting modes that travel along the field 
lines at a velocity $u_f$, which is proportional to the perpendicular 
flow shear, see \eref{u_f}. 
Modes can be excited by both the ion temperature 
gradient and parallel shear flow --- in some cases the drives compete and 
instability 
can be suppressed and in others they 
combine to enhance instability (see \figs{figure4}{figure5}).
We define the Mach number of the moving perturbation as $M= u_f/c_s$ where 
$c_s$ is the sound speed.  
In \sref{secconvection} we show that when $M > 1$, perturbations are swept 
downstream until they are damped away --- thus only transitory growth can take 
place for $M > 1$. 
At very large flow shear the transitory growth is substantial: 
the number of exponentiations before decay is proportional to $\sqrt{M}$, 
see \sref{sectransitory} and equation~(\ref{asymptrans}).

While it would be foolish to conclude too much about the behaviour of fusion 
devices from this simple model, a number of trends stimulate some speculation.
First, one expects that when $u_f$ exceeds the characteristic propagation of 
the instability, exponential growth is suppressed. 
The characteristic propagation obviously depends on the instability.
For example, electron instabilities have propagation speeds of order 
of the electron thermal velocity.
It also seems likely that, for ion temperature gradient driven modes,  
turbulent suppression is maximized at values of $M$ close 
to, or just above, one. 
Increasing the flow shear above this value may lead to the destabilization 
of transient instabilities and strong resulting turbulence. 
Barnes \etal~\cite{barnes} and Highcock \etal~\cite{highcock} have recently shown that this trend is indeed 
seen in gyro-kinetic simulations. 
Since turbulence is suppressed by perpendicular shear flow but enhanced 
by parallel flow we expect the sheared toroidal flow to give more suppression in regions of strong poloidal 
field. 
Indeed simulations by Roach \etal~\cite{roach} confirm this trend. 
This also suggests strong suppression in Spherical tokamaks, 
where the poloidal field often exceeds the toroidal field on the 
outside of the magnetic surfaces (where the curvature drive is destabilizing). 
Work is in progress to establish more clearly the trends in shear 
flow stabilization and to optimize transport.


\ack

The authors wish to thank Ian Abel, Michael Barnes, Paul Dellar, Edmund Highcock, 
Felix Parra, Colin Roach, Alexander Schekochihin and Alessandro Zocco 
for stimulating discussions. We also thank the Leverhulme Trust Network for Magnetized Plasma Turbulence 
for travel support.
This work was supported in part by the United Kingdom Engineering and 
Physical Sciences Research Council under grant EP/G003955, by Funda\c{c}\~ao para a Ci\^encia e 
Tecnologia and by the European Communities under the contracts of Association between EURATOM 
and CCFE and EURATOM and IST. The views and opinions expressed herein do not 
necessarily reflect those of the European Commission.
Simulations were performed at IST Cluster and HPC-FF (Juelich).


\section*{Appendix: Derivation of Moment Equations}

In this appendix we give details of the derivation of the set of moment equations presented in \sref{secresponse}. As discussed there, the electrons are taken to have an isothermal, or Boltzmann, density response, $\delta n_e$, to the perturbing potential, $\phi$:
\be
\delta n_e = n_0 \frac{e \phi}{T_e},
\label{eresponse}
\ee
where $T_e$ is the equilibrium electron temperature. The ion response is obtained from the gyro-kinetic equation~\cite{frieman,sugama,artun_93}. 
The ion distribution function, $f$, correct to first order in the gyro-kinetic expansion in $\omega / \Omega$ is:
\bea
f = F_0\(\epsilon,{\bm R}\) + h\({\bm R},w_\parallel,w_\perp,t\), \\
\delta f = - \frac{e \phi\({\bm r},t\)}{T_i} F_0 + h\({\bm R},w_\parallel,w_\perp,t\),
\label{iresponse}
\eea
where $\delta f$ is the perturbation from equilibrium, the particle energy is $\epsilon = mw^2 / 2 + e \phi\({\bm r},t\)$, the particle velocity is $\bm v$, the velocity variable ${\bm w} = {\bm v} - {\bm V}_0$ and the guiding center position satisfies ${\bm R} = {\bm r} - {\bf b} \times {\bm v}/\Omega \equiv {\bm r} - \bm \rho$. With the ion thermal velocity $\vthi^2 = 2 T_i / m$ for an equlibrium ion temperature $T_i$, the Maxwellian $F_0 = \(n_0/\pi^{3/2}\vthi^3\)\exp\(-\epsilon / T_i\)$.

The distribution function of gyrocenters, $h$, is independent of the gyroangle of the particle motion, $\zeta$, and is defined by:
\bea
\fl \frac{\p h}{\p t} + \( w_\parallel {\bf b} + {\bm V}_0\) \cdot \nabla h + \frac{1}{B_0}\left\{\lang \phi \rang, h \right\} - \lang C^l\(h\)\rang \nonumber \\
\fl = \frac{1}{B_0}\left[\frac{1}{l_n} + \(\frac{\epsilon}{T_i}- \frac{3}{2}\)\frac{1}{l_T} + \frac{8 w_\parallel}{3 c_s l_v}\right] F_0 \frac{\p \lang \phi\rang}{\p y} + \frac{e}{T_i} \left[\frac{\p \lang \phi \rang}{\p t} + {\bm V}_0 \cdot \nabla \lang \phi \rang \right] F_0. 
\label{hieqEdmund}
\eea
The collision operator appearing here, $C^l$, is the linearized part of the self-collision operator~\cite{hsbook} acting on the ion distribution function $F_0 + h$ and is discussed further below. The angled brackets denote the average of the enclosed quantity over the gyroangle at constant $\bm R$: $\lang A \( {\bf r}\)\rang = \(2\pi\)^{-1}\oint A \({\bm R} +  \bm \rho\) d\zeta$ and the Poisson bracket is defined as follows, where the spatial gradient is taken at constant ${\bm w}$:
\bea
\left\{\lang \phi \rang, h \right\} =  \( \nabla \lang \phi \rang \times \nabla h \)\cdot {\bf b}.
\eea
The ion response is driven by the background density and temperature gradients, as well as the parallel flow shear:
\be
\frac{1}{l_n} = \frac{d}{dx}\ln n_0, \qquad
\frac{1}{l_T} = \frac{d}{dx}\ln T_i, \qquad
\frac{1}{l_v} = \frac{1}{L_v}\frac{V_0}{c_s}{\bf \hat e_v} \cdot {\bf \hat{z}}.
\ee
The ion sound speed, $c_s$, is defined in \sref{secresponse}. The double shearing coordinate transformation discussed in \sref{secgeometry} is now implemented, with $u_f = V_0 \(l_s/L_v\){\bf \hat{e}_v} \cdot {\bf \hat{y}}$, so:
\bea
\fl x^\prime = x, \qquad
y^\prime = y - \frac{x}{l_s}\(z + u_f t\), \qquad
z^\prime = z + u_f t, \qquad
t^\prime = t, \\
\fl \frac{\p }{\p x} = \frac{\p }{\p x^\prime} - \frac{z^\prime}{l_s}\frac{\p }{\p y^\prime} \qquad
\frac{\p }{\p y} = \frac{\p }{\p y^\prime}, \qquad
\frac{\p }{\p z} = \frac{\p }{\p z^\prime} - \frac{x}{l_s}\frac{\p }{\p y^\prime}, \nonumber \\
\fl \frac{\p}{\p t} = \frac{\p}{\p t^\prime} + u_f \frac{\p}{\p z^\prime} - u_f \frac{x}{l_s}\frac{\p}{\p y^\prime} = \frac{\p}{\p t^\prime} + u_f \frac{\p}{\p z^\prime} - {\bm V}_0  \cdot \nabla + \ldots .
\eea
Leading order corrections in $x/l_s$ will be retained only when multiplied by $k_y$. That is, once the twisting nature of the perturbed structures arising due to the magnetic field shear has been accounted for by the frame transformation, only the leading effect of the background gradients, their drive of drift waves, is retained. The equation for $h$ takes the form:
\bea
\fl \frac{\p h}{\p t} + \( w_\parallel + u_f\) \frac{\p h}{\p z^\prime} + \frac{1}{B_0}\left\{\lang \phi \rang, h \right\} - \lang C^l\(h\)\rang \nonumber \\
\fl = \frac{1}{B_0}\left[\frac{1}{l_n} + \(\frac{\epsilon}{T_i}- \frac{3}{2}\)\frac{1}{l_T} + \frac{8 w_\parallel}{3 c_s l_v}\right] F_0 \frac{\p \lang \phi\rang}{\p y^\prime} + \frac{e}{T_i} \left[\frac{\p \lang \phi \rang}{\p t} + u_f \frac{\p \lang \phi \rang}{\p z^\prime}  \right] F_0.
\label{hieq}
\eea
The gyroaverages and Poisson bracket are now taken to represent their transformed values. The primes on the transformed variables are dropped from hereon.

For clarity in identifying the effects of the flow shear, we consider the collisional limit of this system and so introduce the subsidiary ordering:
\be
\nu~\gg~\omega,~\omega^*,~\vthi k_\parallel,~u_f k_\parallel,~\nu k^2 \rho^2.
\label{suborderapp}
\ee
Here $\nu$ is the ion-ion collision frequency, electron mass corrections are neglected, $k\rho \sim \mathcal O \(\sqrt{\omega/\nu}\)\ll 1$ where the ion gyroradius is $\rho = \vthi /\Omega$ and $\omega^*$ represents the drift frequencies associated with the background gradients, which are identified in \sref{seclinsys}. The distribution function is expanded in $\omega/\nu$ such that: $h = h^{(0)} + h^{(1)} + \ldots$, where the order of expansion is denoted by a superscript. We then proceed with the solution of the gyro-kinetic equation \eref{hieq} order by order, taking all quantities on the right of \eref{suborderapp} to be of the same order.

Care must be taken when considering the ion self-collision operator. The Landau form, acting on a distribution function $f\({\bm r},{\bm v},t\)$, is~\cite{hsbook}:
\be
\fl C\(f,f\) = \nu_i \left.\frac{\p}{\p \bm v}\right|_{\bm r} \cdot \int d^3 v^\prime {\sf U} \cdot \left[ f\(\bm v ^\prime\) \left.\frac{\p}{\p \bm v}\right|_{\bm r} f \(\bm v\) - f\(\bm v\) \left.\frac{\p}{\p \bm v ^\prime}\right|_{\bm r} f \(\bm v ^\prime\)\right],
\label{colop}
\ee
where the relative velocity $\bm u = \bm v - \bm v^\prime$, the tensor ${\sf U} = \(u^2 {\sf I} - {\bm u}{\bm u}\)/u^3$, $\sf I$ is the identity tensor, $\nu_i = e^4 \ln \Lambda/8 \pi \epsilon_0^2 m^2$ and $\ln \Lambda$ is the Coulomb logarithm. The collision operator is Galilean invariant and for a Maxwellian, $F_M\(v\) = \(n_0/\pi^{3/2}\vthi^3\)\exp\(-v^2 / \vthi^2 \)$, $C\(F_M,F_M\) = 0$. Therefore the linearized collision operator appearing in \eref{hieq} is:
\be
C^l\(h\) \equiv C\(h,F_M\) + C\(F_M,h\).
\label{lincolop}
\ee
As discussed by Abel \etal~\cite{abel}, the velocity derivative appearing in the collision operator is taken at constant particle position, $\bm r$, whilst the gyroaverage in \eref{hieq} is taken at constant guiding centre position, ${\bm R}$. As $F_M$ does not vary on the scale of the gyroradius, this may be conveniently dealt with by using the Fourier representation of $h$, which is a function of guiding centre position, within the collision term:
\be
h\({\bf R}\) = \sum_{\bf k} e^{i{\bf k} \cdot {\bf R}}h_k = \sum_{\bf k} e^{i{\bf k} \cdot \({\bf r} -  {\bm \rho}\)}h_k.
\ee
Then, as $k \rho \ll 1$:
\bea
\fl \lang C^l\(h\)\rang_{\bm R} & = \sum_{\bm k} \lang e^{i{\bm k} \cdot {\bm r}} C^l\(h_k e^{-i{\bm k} \cdot {\bm \rho}} \) \rang_{\bm R} \nonumber \\
\fl & = \sum_{\bm k} e^{i{\bf k} \cdot {\bm R}} \lang e^{i{\bm k} \cdot {\bm \rho}} C^l\(h_k e^{-i{\bm k} \cdot {\bm \rho}} \) \rang_{\bm R} \nonumber \\
\fl & \approx  \sum_{\bm k} e^{i{\bm k} \cdot {\bm R}} \lang \(1 + i{\bm k}\cdot {\bm \rho} + \ldots\) C^l\(h_k \(1 - i{\bm k}\cdot {\bm \rho} + \ldots \)\) \rang_{\bm R}.
\eea
Note that the gyroaverage eliminates terms of $\mathcal O \(k \rho\)$ in this expression.

Now, to lowest order, \eref{hieq} gives:
\be
C^l\(h^{(0)}\) = 0.
\ee
This indicates~\cite{hmbook} that $h^{(0)}$ has the form of a perturbed Maxwellian in the guiding center variable $\bm R$, with the perturbed density, $\delta n$, temperature, $\delta T$ and parallel velocity, $\delta V_\parallel$:
\be h^{(0)} = \left[\frac{\delta n}{n_0} + \frac{\delta T}{T_i}\(\frac{\epsilon}{T_i} - \frac{3}{2}\) + \frac{2 w_\parallel \delta V_\parallel}{\vthi^2} \right] F_M\(w, {\bm R}\). \label{h0}\ee

With all quantities evaluated at the guiding center position and the collisional contribution denoted as $C^{(1)}$, the next terms in the expansion of \eref{hieq} give:
\bea
\fl \frac{\p h^{(0)}}{\p t} + \( w_\parallel + u_f\) \frac{\p h^{(0)}}{\p z} + \frac{1}{B_0}\left\{ \phi, h^{(0)} \right\} - C^{(1)} \nonumber \\
\fl = \frac{1}{B_0}\left[\frac{1}{l_n} + \(\frac{\epsilon}{T_i}- \frac{3}{2}\)\frac{1}{l_T} + \frac{8 w_\parallel}{3 c_s l_v }\right] F_M \frac{\p \phi}{\p y} + \frac{e}{T_i} \(\frac{\p \phi }{\p t} + u_f \frac{\p \phi}{\p z} \) F_M.
\label{hieqexp}
\eea
The density, parallel velocity and energy moments of \eref{hieqexp} give a set of equations describing the evolution of the parameters $\delta n$, $\delta T$ and $\delta V_\parallel$ of the perturbed Maxwellian \eref{h0}. The collision term, $C^{(1)}$, has three contributions. The first, proportional to $\({\bm k}\cdot {\bm \rho}\)^2 C^l\(h^{(0)}\)$, is zero, given the form of $h^{(0)}$. The second arises from terms of the form: $C^l\(h_k \({\bm k}\cdot {\bm \rho}\)^2\)$ and will not contribute to the moment equations due to the conservation properties of the collision operator: $\int \(1,w_\parallel,w^2\) C^{l}\(g\) d{\bm w} = 0$, where $g$ is an arbitrary function. The remaining contribution is:
\be
\sum_{\bm k} e^{i{\bm k} \cdot {\bm R}} \lang {\bm k}\cdot {\bm \rho}~C^l\(h_k {\bm k}\cdot {\bm \rho} \) \rang_{\bm R}.
\label{eqc1defn}
\ee
Defining:
\be
\varphi \({\bm w}\) = - h_k\({\bm w}\) {\bm k}\cdot {\bm \rho} = h_k\({\bm w}\) {\bm w}\cdot {\bm \sigma},
\ee
where ${\bm \sigma} = {\bf b} \times {\bm k} / \Omega$ and using ${\bm u}\cdot {\sf U} = {\sf U} \cdot {\bm u} = 0$, the definition \eref{colop} shows that \eref{eqc1defn} reduces to:
\be
\sum_{\bm k} e^{i{\bm k} \cdot {\bm R}} \lang {\bm w} \cdot {\bm \sigma} \nu_i \frac{\p}{\p \bm w} \cdot \int d^3 w^\prime {\sf U} \cdot \left[\frac{\p}{\p \bm w} \varphi\({\bm w}\) - \frac{\p}{\p {\bm w}^\prime}\varphi\({\bm w}^\prime\)\right] \rang_{\bm R},
\ee
with:
\be
\fl \left[\frac{\p}{\p \bm w} \varphi\({\bm w}\) - \frac{\p}{\p {\bm w}^\prime}\varphi\({\bm w}^\prime\)\right] = \left( {\bm s} {\bm \sigma} \cdot {\bm u} + {\bm \sigma} {\bf u}\cdot {\bm s}\right) \frac{\delta T_k}{\vthi^2 T_i} + \left( {\bm b} {\bm u}\cdot {\bm \sigma} +  {\bm \sigma} {\bm u} \cdot {\bm b} \right) \frac{ 2 \delta V_{\parallel k}}{\vthi^2},
\label{eqc1}
\ee
where we have used the sum and difference velocity variables: ${\bm s} = {\bm w} + {\bm w}^\prime$ and ${\bm u} = {\bm w} - {\bm w}^\prime$. If we weight \eref{eqc1} by a function of velocity $g\({w_\parallel, w_\perp}\)$ and integrate over velocity space, integrating by parts noting that $f_M \rightarrow 0$ as $\left|{\bm w}\right| \rightarrow 0$, gives the resulting integral, $M_C$, to be:
\bea
\fl M_C = - \frac{{\bm \sigma}}{16} \cdot \int d^3s \int d^3u \left[{\sf I}\(g - g^\prime\) + {\bm w}\frac{\p{g}}{\p{\bm w}} - {\bf w}^\prime \frac{{\p{g^\prime}}}{\p{{\bm w}^\prime}}\right]\cdot{\sf U} \nonumber \\
\fl \cdot \left[\left( {\bm s} {\bm \sigma} \cdot {\bm u} + {\bm \sigma} {\bm u}\cdot {\bm s}\right) \frac{\delta T_k}{\vthi^2 T_i} + \left( {\bm b} {\bm u}\cdot {\bm \sigma} +  {\bm \sigma} {\bm u} \cdot {\bm b} \right) \frac{ 2 \delta V_{\parallel k}}{\vthi^2}\right] F_M\(\frac{u}{\sqrt{2}}\)F_M\(\frac{s}{\sqrt{2}}\),
\label{eqmc}
\eea
where $g^\prime = g\(w_\parallel^\prime, w_\perp^\prime \)$.

Now we take the density, parallel velocity and energy moments of \eref{hieqexp}, which correspond to $g = \left\{1, w_\parallel, \(\vthi^{-2} w^2 - 3/2\)\right\}$. For $g=1$, the collisional contribution is seen to be zero by \eref{eqmc}, as required for particle conservation. To evaluate the collisional dissipation of the parallel velocity, note that the term in \eref{eqmc} proportional to $\delta T$ is odd in ${\bm s}$ for $g = w_\parallel$. So the dissipation is proportional to $\delta V_\parallel$:
\bea
\int d^3w w_\parallel C^{\(1\)} &= - \frac{4 \sqrt{2}}{5} \frac{\nu_i n_0^2}{\sqrt{\pi} \vthi \Omega^2} \sum_{\bm k} e^{i {\bm k}\cdot {\bm R}} k_\perp^2 \delta V_{\parallel k} \nonumber \\
& \equiv n_0 \nu \nabla_\perp^2 \delta V_\parallel .
\eea
Similarly, the dissipation arising in the energy moment only retains a contribution from the term proportional to $\delta T$ in \eref{eqmc}:
\bea
\int d^3w \(\frac{w^2}{\vthi^2} - \frac{3}{2}\) C^{\(1\)} & = - \frac{4 \sqrt{2}}{3} \frac{\nu_i n_0^2}{\sqrt{\pi} \vthi \Omega^2 } \sum_{\bm k} e^{i {\bm k}\cdot {\bm R}} k_\perp^2 \frac{\delta T_{k}}{T_i} \nonumber \\
& \equiv \frac{3}{2} \frac{n_0}{T_i} \chi \nabla_\perp^2 \delta T .
\eea
Hence, the evolution equations of the perturbed variables are:
\bea
\fl \(\frac{\p}{\p t} + u_f \frac{\p}{\p z}\) \frac{\delta n}{n_0} + \frac{\p}{\p z}\delta V_\parallel + \frac{1}{B_0}\left\{\phi, \frac{\delta n}{n_0} \right\} = \frac{1}{l_n B_0}\frac{\p \phi}{\p y} + \frac{e}{T_i}\(\frac{\p}{\p t} + u_f \frac{\p}{\p z}\) \phi,
\label{conteqn} \\
\fl \(\frac{\p}{\p t} + u_f \frac{\p}{\p z}\) \delta V_\parallel + \frac{1}{m}\frac{\p}{\p z}\(T_i \frac{\delta n}{n_0} + \delta T\) + \frac{1}{B_0}\left\{ \phi, \delta V_\parallel \right\} = \frac{c_s}{l_v B_0} \frac{\p \phi}{\p y} + \nu \nabla_\perp^2 \delta V_\parallel,
\label{vpareqn} \\
\fl \(\frac{\p}{\p t} + u_f \frac{\p}{\p z}\) \frac{\delta T}{T_i} + \frac{2}{3} \frac{\p}{\p z}\delta V_\parallel + \frac{1}{B_0}\left\{ \phi, \frac{\delta T}{T_i} \right\} = \frac{ n_0}{l_T B_0}\frac{\p \phi}{\p y} + \chi \nabla_\perp^2 \frac{\delta T}{T_i},
\label{energyeqn}
\eea
where:
\be 
\left( \nu,~\chi \right) = \left(\frac{4}{5},~\frac{8}{9}\right) \sqrt{\frac{2}{\pi}}\frac{\nu_i n_0}{\vthi \Omega^2}.
\ee

For simplicity we now assume that the equilibrium ion and electron temperatures are equal: $T_{i} = T_{e} = T$. Upon imposing quasineutrality with the isothermal electron response given in \eref{eresponse}, the electrostatic potential may be eliminated from these moment equations in favour of the perturbed density:
\be
\frac{\delta n_e}{n_0} = \frac{e\phi}{T} = \frac{\delta n_i}{n_0} = \(\frac{\delta n}{n_0} - \frac{e \phi}{T}\).
\ee
This leaves a closed system for the evolution of the perturbed fluid variables: $\delta n$, $\delta V_\parallel$ and $\delta T$. We introduce the following normalizations, which are widely used in gyro-kinetic codes~\cite{GS2}:
\bea
x = \rho_s \tilde x, \qquad
& y =\rho_s \tilde y, \qquad
& z = l_s \tilde z, \qquad
t = \frac{l_s}{c_s} \tilde t, \\
\tV = \frac{\delta V_\parallel}{c_s}\frac{l_s}{\rho_s}, \qquad
& \tT = \frac{\delta T}{T}\frac{l_s}{\rho_s}, \qquad
& \tn = \frac{\delta n}{n_0}\frac{l_s}{\rho_s},
\eea
where $\rho_s = c_s/\Omega$ is the sound Larmor radius. We also introduce an effective Mach number, associated with the convective effect of the perpendicular flow shear, defined as:
\be
M=\frac{u_f}{c_s}.
\ee
Dropping the tilde denoting the final coordinate transformation for convenience, the set of equations describing the evolution of a perturbation of the system finally takes the form:
\bea
\fl \(\frac{\p \tn}{\p t}+M\frac{\p \tn}{\p z}\)
+\frac{\p \tV}{\p z}=\frac{3}{8}\frac{l_s}{l_n}\frac{\p \tn}{\p y},\\
\fl \(\frac{\p \tV}{\p t}+M\frac{\p \tV}{\p z}\)
+\frac{3}{8}\frac{\p}{\p z}\(2\tn+\tT\)+ \frac{3}{8}\left\{\tn,\tV \right\} =
\frac{3}{8}\frac{l_s}{l_v}\frac{\p\tn}{\p y}
+\tilde\nu_\perp\nabla_\perp^2\tV,\\
\fl \(\frac{\p \tT}{\p t}+M\frac{\p \tT}{\p z}\)
+\frac{2}{3}\frac{\p \tV}{\p z} + \frac{3}{8}\left\{\tn,\tT\right\}
=\frac{3}{8}\frac{l_s}{l_T}\frac{\p\tn}{\p y}
+\tilde\chi_\perp\nabla_\perp^2\tT,
\eea
where
\be
\nabla_\perp^2=\frac{\p^2}{\p y^2}+
\(\frac{\p}{\p x}-z\frac{\p}{\p y}\)^2
\ee
and
\bea
\left(\tilde\nu_\perp, \tilde\chi_\perp \right) & = \left( \frac{\nu}{\(c_s \rho_s^2/l_s\)},\frac{\chi}{\(c_s \rho_s^2/l_s\)}\right) \nonumber \\
& = \left(\frac{9}{40},\frac{1}{4}\right)\sqrt{\frac{2}{3}}\frac{l_s n_0 e^4 \ln \Lambda}{8 \pi^{3/2} \epsilon_0^2T^2}.
\eea



\end{document}